# Formalising Sample Size Calculations for the Development of Risk Prediction Models: The Importance of Accounting for Performance Variability

Menelaos Pavlou[1*], Rumana Z. Omar[1], Gareth Ambler[1]

[1] Department of Statistical Science, UCL, UK;

[*] Corresponding author

**Abstract**

Sample size calculations for developing prediction models typically aim to ensure that the expected value of a performance measure meets a prespecified target. For example, a key measure is the calibration slope (CS) which quantifies model overfitting; the sample size is often chosen so that the expected CS equals 0.9, close to the ideal value of 1. However, because of sampling variability, model performance can vary substantially across development samples of the recommended size. When variability is high, the probability of obtaining a model with performance close to the target may be unacceptably low.

We propose a framework which enables sample size calculations that incorporate both the expected value and the variability for a given performance measure. The framework is illustrated for binary outcomes and logistic regression but applies to other outcomes and model types. To explicitly account for variability, we introduce the probability of acceptable performance (PrAP). For example, a model may be considered acceptable if the CS lies within a prespecified range (e.g., 0.85 to 1.15) and the sample size might be chosen to ensure that PrAP exceeds some target (e.g., 80%). Under existing approaches PrAP can be low, especially when the specified number of predictors is small, which can also translate into large variability for individual predicted probabilities. The use of shrinkage tends to improve PrAP.

Our findings highlight the importance of accounting for performance variability to ensure robust model development.

## 1 Introduction

Risk prediction models can estimate the probability of a patient experiencing a health event (e.g. in-hospital death following cardiac surgery)[1] based on individual patient characteristics (e.g. age, comorbidities, imaging results). As they provide individualised predictions, they have the potential to support decision-making by clinicians and patients alike.

Given a list of candidate predictor variables identified from literature reviews and/or expert opinion, a prediction model can be developed by estimating the association between predictors and the outcome. The dataset used for model fitting is called the development or training dataset. The association can be estimated using regression (e.g. logistic and Cox regression for binary and survival outcomes, respectively) or other methods, including machine learning. In this work we primarily focus on regression-based models for binary outcomes fitted with Maximum Likelihood Estimation (MLE) although most of the ideas and definitions in Section 2 apply to other types of outcomes, classes of models and estimation methods.

Before a model can be used in practice, its predictive performance needs to be assessed, i.e. the model should be validated in a new dataset, the validation dataset, by calculating measures of predictive performance [2, 3]. Commonly used measures include the calibration slope and calibration in-the-large to assess the agreement between observed and predicted probabilities, the C-statistic for discrimination, the Brier Score for overall predictive accuracy and net benefit for clinical utility[4].

The sample size of the development dataset plays a critical role in determining the model's predictive accuracy in new patients. A sample size that is too small relative to the number of predictors or regression parameters may lead to model overfitting, meaning that the model may not generalise well to new patients. Model overfitting is typically quantified by the calibration slope (CS) which assesses the agreement between the observed and predicted probabilities in validation data. A perfectly calibrated model has CS of 1 while values much smaller than 1 (e.g. <0.85) can be indicative of substantial overfitting. Larger degrees of overfitting also suggest greater performance losses in terms of other performance measures.

Currently used sample size calculations for the development of prediction models aim to ensure that the 'expected performance' of the model will meet a prespecified target for a given performance measure [5, 6]. For example, current approaches for the CS use either analytical formulae[5] or simulation[7] to determine the sample size such that the CS meets a prespecified target value on average. A commonly used target is 0.9. This means, if one were to collect development datasets of the recommended size, fit the model to each and validate all models on independent datasets from the same population, then the average CS across those datasets would be 0.9. In practice, due to sampling variability, model performance can vary considerably across development samples. As a result, even if we adhere to the recommended sample size, a high variability in performance means the probability of obtaining a model with performance close to the target may be unacceptably low. In addition to aggregate performance measures, sample size calculations have more recently been considered for individual predicted probabilities.[8]

In this work we propose an adaptation of sample size calculations that explicitly accounts for variability in performance. Instead of aiming for performance on average, our proposed adaptation targets the probability of obtaining a model with acceptable performance for a chosen measure. The structure of the paper is as follows. In Section 2 we formally define the potential targets of sample size calculations for the development of prediction models. Section 3 discusses limitations of existing approaches for sample size calculations and briefly clarifies assumptions required for their validity in practice. In Section 4, we provide a heuristic bias-correction adjustment for an existing formula targeting the expected calibration slope. In Section 5 we demonstrate the importance of accounting for variability in performance and introduce our proposed adaptation, which is discussed in detail for the CS. We also discuss a computationally efficient analytical calculation for the new method. Section 6 investigates the potential usefulness of a simple linear shrinkage factor approach for improving model calibration in datasets whose size is based on existing and newly proposed approaches. Section 7 presents an application of the methods to a real cardiac dataset. We conclude with a Discussion.

## 2 Formalising sample size calculations for the development of prediction models

In this section we introduce the concepts underlying sample size calculations for prediction model development, including the data-generating mechanism that gives rise to outcomes and predictors, the chosen prediction model fitted to a finite development dataset, and measures of predictive performance evaluated in new data.

Let $P(Y, \boldsymbol{X}; \boldsymbol{\varphi})$ denote the joint distribution of an outcome $Y$ and a $p$-dimensional vector of covariates $\boldsymbol{X} = \left(X_1, X_2, \ldots, X_p\right)^\top$, where $\boldsymbol{\varphi}$ represents the parameters of this distribution. In prediction modelling, the covariates $\boldsymbol{X}$ represent candidate predictors, identified from the literature and expert knowledge. The joint distribution can be factorised as $P(Y, \boldsymbol{X}; \boldsymbol{\varphi}) = P(Y|\boldsymbol{X}; \boldsymbol{\varphi}_1)P(\boldsymbol{X}; \boldsymbol{\varphi}_2)$ where $\boldsymbol{\varphi}_1$ and $\boldsymbol{\varphi}_2$ parameterise the conditional and marginal distributions, respectively. Together, these components define the data generating mechanism (DGM) that gives rise to the observed data. The true DGM is typically unknown in practice. In prediction modelling, the primary interest lies in the conditional distribution $P(Y|\boldsymbol{X})$. While the marginal distribution of the covariates, $P(\boldsymbol{X})$, is not explicitly modelled when fitting a model, it can become relevant in the context of sample size calculations, as discussed later.

*The chosen model, M, for modelling $P(Y|\boldsymbol{X})$*

A variety of modelling approaches are available for estimating $P(Y|\boldsymbol{X})$. In clinical prediction modelling, regression-based methods are most commonly used. In particular, linear, logistic and Cox regression are routinely used for continuous, binary and time-to-event outcomes, respectively. These models can be fitted using MLE, or penalised methods, such as Ridge and LASSO. Alternative approaches include machine learning techniques such as random forests, gradient boosting machines, neural networks and others[9, 10]. In this work we primarily consider regression-based models and focus on binary outcomes modelled with a logistic model. Nevertheless, the framework is presented in general terms, and the underlying principles apply to other outcome types and modelling approaches.

*Example*. For a binary outcome $Y$, the chosen model is often logistic regression of the form

$$\text{logit}\big(P(Y|\boldsymbol{X};\boldsymbol{\beta})\big) = \beta_0 + \boldsymbol{X}^\top \boldsymbol{\beta}_1, \tag{1}$$

where $\boldsymbol{\beta} = (\beta_0, \boldsymbol{\beta}_1^\top)^\top$ are regression parameters. We note that the chosen model $M$ need not be correctly specified; that is, the true conditional distribution $P(Y|\boldsymbol{X};\boldsymbol{\varphi}_1)$ may not coincide with the chosen model form (1). Also, the chosen model form in the case of regression need not be linear in the predictor effects; it may instead incorporate more flexible structures to accommodate non-linear effects and interactions.

*Development datasets*

Let $\mathcal{D}^n = \{(y_i, \boldsymbol{x}_i)\}_{i=1}^n$ denote a random sample of size $n$ drawn from $P(Y, \boldsymbol{X}; \boldsymbol{\varphi})$, referred to as the development (or training) dataset. For a chosen model $M$ and fitting procedure, the parameters for modelling $Y|\boldsymbol{X}$ are estimated by fitting the model to $\mathcal{D}^n$.

*Example*. For a binary outcome $Y$, the logistic model (1) is fitted to $\mathcal{D}^n$ using MLE to obtain estimates of $\boldsymbol{\beta}$. We denote these estimates by $\widehat{\boldsymbol{\beta}}_n = \big(\hat{\beta}_{0,n}, \widehat{\boldsymbol{\beta}}_{1,n}^\top\big)^\top$. Under standard regularity conditions $\widehat{\boldsymbol{\beta}}_n$ is consistent and asymptotically normal.

*Validation datasets*

To evaluate predictive performance, we consider an independent validation (or test) dataset $\mathcal{V}^m = \big\{\big(y_{\text{val},i}, \boldsymbol{x}_{\text{val}.i}\big)\big\}_{i=1}^m$ which is a random sample of size $m$ drawn from the same population as the training data. For a model fitted on $\mathcal{D}^n$, the predicted probability for a new observation with covariate values $\boldsymbol{x}^\star$ is denoted by $\hat{\pi}_n(\boldsymbol{x}^*)$. The vector of predicted probabilities for the validation data is $\widehat{\boldsymbol{\pi}}_n(\mathcal{V}^m) = \big(\hat{\pi}_n(\boldsymbol{x}_{\text{val},1}), \dots, \hat{\pi}_n(\boldsymbol{x}_{\text{val},m})\big)^\top = \big(\hat{\pi}_{1,n}, \dots, \hat{\pi}_{m,n}\big)^\top$, where subscript $n$, emphasises that predictions are based on a fitted model using a training sample of size $n$.

*Example*. For model (1), the individual predicted probability (IPP) for a new observation is $\hat{\pi}_n(\boldsymbol{x}^*) = \text{logit}^{-1}\big(\hat{\beta}_{0,n} + \boldsymbol{x}^{*\top}\, \widehat{\boldsymbol{\beta}}_{1,n}\big)$. For the validation dataset $\mathcal{V}^m$, $\hat{\pi}_{i,n} = \text{logit}^{-1}\big(\hat{\beta}_{0,n} + \boldsymbol{x}_{\mathbf{val},\boldsymbol{i}}^\top \widehat{\boldsymbol{\beta}}_{1,n}\big), i = 1, \dots m$.

*Measures of predictive performance* $\boldsymbol{\theta}$

The predictive performance of a model fitted on $\mathcal{D}^n$ is assessed on $\mathcal{V}^m$ using a performance measure, $\theta$, defined as a function of the observed outcomes and predicted outcomes/probabilities in the validation data. Common examples **include** the calibration slope, the C-statistic and the Brier

Score. We let $\hat{\theta}(\hat{\boldsymbol{\pi}}_n, \mathcal{V}^m)$ denote the estimate of $\theta$ computed on the validation dataset. For notational simplicity we write this as $\hat{\theta}_{n,m}$.

For a model fitted to a specific training dataset of size $n$, the corresponding 'true performance' can be defined as the limiting value obtained with an infinitely large validation dataset i.e.,

$$\hat{\theta}_n = \lim_{m \to \infty} \hat{\theta}(\hat{\boldsymbol{\pi}}_n, \mathcal{V}^m) \,.$$

In this work we focus on the sample size requirements for model development. Hence, we are interested in evaluating the true performance conditional on the development sample size, $\hat{\theta}_n$, rather than the estimated value $\hat{\theta}_{n,m}$ which is subject to noise due to the validation data being of finite size. Hereafter, we refer to $\hat{\theta}_n$ as the model's performance.

*Optimal value for a performance measure under the chosen model*

We define the optimal value of the performance measure $\theta$, denoted as $\theta_{\text{opt}}$, as the value obtained when the chosen model *M* is fitted to an infinitely large development dataset, and evaluated on an infinitely large validation dataset, both drawn from $P(Y, \boldsymbol{X}; \boldsymbol{\varphi})$:

$$\theta_{\text{opt}} = \lim_{n \to \infty} \hat{\theta}_n \,.$$

In simulation studies, the true outcome generating mechanism $P(Y | \boldsymbol{X}; \boldsymbol{\varphi}_1)$ is known. When the assumed model $M$ coincides with the true model (e.g., logistic regression of the form (1)), the optimal value can be computed using predicted probabilities/outcomes evaluated at the true parameter values.

*Examples.* For the calibration slope, $s$, the optimal value is $s_{\text{opt}} = 1$. For the C-statistic, the optimal value depends on the predictive information contained in the covariates, and the modelling approach. For example, if the chosen modelling approach is logistic regression with linear predictor effects, the optimal C-statistic may be lower than that obtained using more flexible models (e.g., random forests) if the true relationship between $Y$ and $\boldsymbol{X}$ involves non-linearities or interactions that cannot be fully captured by the chosen model. As a measure of overall predictive accuracy, the Mean Absolute Prediction Error (MAPE), defined as the expected absolute difference between the true and predicted outcomes, is only calculable in simulation settings. For logistic regression, if the assumed model coincides with the true model, then $\text{MAPE}_{\text{opt}} = 0$.

*Sampling distribution of $\hat{\theta}_n$ and $\hat{\pi}_n(\boldsymbol{x}^*)$*

The sampling distribution of $\hat{\theta}_n$ and $\hat{\pi}_n(\boldsymbol{x}^*)$ is obtained by repeatedly drawing training datasets $\mathcal{D}^n$ of size $n$ from $P(Y, \boldsymbol{X}; \boldsymbol{\varphi})$, fitting the chosen model to each dataset and either calculating the individual predicted probability for $\boldsymbol{x}^*$ or evaluating performance on large independent validation datasets drawn from the same population. Because $\mathcal{D}^n$ is random, $\hat{\theta}_n$ and $\hat{\pi}_n(\boldsymbol{x}^*)$ are random variables with expectations $\mathbb{E}(\hat{\theta}_n)$ and $\mathbb{E}(\hat{\pi}_n(\boldsymbol{x}^*))$, and variances $\text{Var}(\hat{\theta}_n)$ and $\text{Var}(\hat{\pi}_n(\boldsymbol{x}^*))$, respectively, where expectations are taken over the distribution of $\mathcal{D}^n$.

*Probability of acceptable performance*

To explicitly account for performance variability, we may define the probability of acceptable performance (PrAP). For a predefined range $(l_\theta, u_\theta)$,

$$\text{PrAP}(\hat{\theta}_n) = \Pr(l_\theta \leq \hat{\theta}_n \leq u_\theta),$$

is the probability that a model developed with sample size $n$ achieves aggregate performance within a desired target range. Sample size calculations can then be formulated based on choosing $n$ so that PrAP is sufficiently high. For example, for the calibration slope we may define $\text{PrAP}(\hat{s}_n) = \Pr(0.85 \leq \hat{s}_n \leq 1.15)$. This approach explicitly incorporates the expected value and the variability of the performance measure.

Analogously, acceptable performance may be defined with respect to an individual predicted probability: $\text{PrAP}(\hat{\pi}_n(\boldsymbol{x}^*)) = \Pr(l_\pi \leq \hat{\pi}_n(\boldsymbol{x}^*) \leq u_\pi)$, for a suitably defined range $(l_\pi, u_\pi)$. For instance, suppose that a clinical intervention is designed to be applied to patients with predicted probability greater than a threshold value $t$. Then, for an individual with covariate values $\boldsymbol{x}^{(1)}$ and true predicted probability $\pi(\boldsymbol{x}^{(1)}) < t$, we may want to ensure that $\Pr(\hat{\pi}_n(\boldsymbol{x}^{(1)}) > t)$ is low.

*Targets for sample size calculations*

The quantities relevant to sample size calculations for developing prediction models are summarised in Table 1. Sample size calculations may target aggregate performance measures $\theta$ or individual predictions $\pi(\boldsymbol{x}^*)$ for a covariate vector $\boldsymbol{x}^*$. Conditional on a development sample of size $n$, quantities of interest may include:

i) Expected Performance: $\mathbb{E}(\hat{\theta}_n)$ and $\mathbb{E}(\hat{\pi}_n(\boldsymbol{x}^*))$.

ii) Variability: $\text{Var}(\hat{\theta}_n)$ and $\text{Var}(\hat{\pi}_n(\boldsymbol{x}^*))$

iii) Probability of acceptable performance (PrAP) for performance measures and individual predictions, respectively: $\text{PrAP}(\hat{\theta}_n) = \Pr(l_\theta \leq \hat{\theta}_n < u_\theta)$ and $\text{PrAP}(\hat{\pi}_n(\boldsymbol{x}^*)) = \Pr(l_\pi \leq \hat{\pi}_n(\boldsymbol{x}^*) \leq u_\pi)$, where $(l_\theta, u_\theta)$ and $(l_\pi, u_\pi)$ define the corresponding acceptable ranges.

We note that in the case of skewed sampling distributions or the presence of outliers, the median (e.g. $\text{Med}(\hat{\theta}_n)$, may be used instead as a measure of central tendency. When sample sizes are reasonably large, the sampling distributions of the performance measures tend to be approximately symmetric and hence $\mathbb{E}(\hat{\theta}_n) \approx \text{Med}(\hat{\theta}_n)$. However, for pragmatic reasons, when we later refer to simulation-based sample size calculations or report simulation results, we use the median rather than the mean to avoid the effect of extreme outliers. For variability, suitable central intervals based on quantiles of the corresponding sampling distribution can be used, e.g. $\left(Q_{0.25}(\hat{\theta}_n), Q_{0.75}(\hat{\theta}_n)\right)$, where $Q_\alpha(\hat{\theta}_n)$ denotes the $a-$quantile of the distribution of $\hat{\theta}_n$.

[Table 1]

# 3 Existing approaches to sample size calculation and their assumptions

## 3.1 Analytical sample size calculations

A natural target for sample size calculations is $\mathbb{E}(\hat{\theta}_n)$[5]. In analytical calculations the training sample size, $n,$ is chosen such that the expected performance of models trained on samples of size $n$, and evaluated on data from the same population, is 'close enough' to $\theta_{\text{opt}}$ (or $\theta_{\text{opt}}$ for simplicity of notation). For example, for the calibration slope $(s)$, $s_{\text{opt}} = 1$; a commonly used target value is $\mathbb{E}(\hat{s}_n) = 0.9$. For discrimination, if $C_{\text{opt}} = 0.7$ we may target a $C$ that lies within 0.02 of the optimal value i.e. $\mathbb{E}(\hat{C}_n) = 0.68$. For overall accuracy, measured for example by MAPE, if the model is correctly specified, then $\text{MAPE}_{\text{opt}} = 0$, and a small target value, e.g., 0.05 might be specified [6].

The seminal work of Riley et al. (2019)[5, 11] derived analytical formulae for regression models fitted with MLE targeting $\mathbb{E}(\hat{\theta}_n)$, for performance measures including the calibration slope and Nagelkerke's $R^2$. These formulae require anticipated inputs such as outcome prevalence, model strength, and the number of predictors, but do not require specification of the joint distribution $P(Y, \boldsymbol{X}; \boldsymbol{\varphi})$.

Analytical approaches are simple and widely used but have limitations. First, they are restricted to regression models fitted by MLE and to a limited set of performance metrics. Second, the formula based on model overfitting, which most often results in the largest sample size, has been shown to substantially underestimate the required sample size for models with high predictive strength (e.g. $C \geq 0.8$) [7]. In Section 4 we provide a heuristic bias-correction adjustment to address this limitation. More fundamentally, analytical approaches focus solely on expected performance and do not consider variability, $\text{Var}(\hat{\theta}_n)$. As noted in the literature[12-15], average performance alone is not a sufficient indicator of model performance; even if $\mathbb{E}(\hat{\theta}_n)$ is close to the target, high variability may mean that acceptable performance cannot be assured for individual datasets. Large values of $\text{Var}(\hat{\theta}_n)$ therefore indicate model instability. Consequently, both the expected performance and variability should be considered when calculating the required sample size for model development. The importance of accounting for variability in performance is demonstrated in Section 5.

## 3.2 Simulation-based sample size calculations

Alternatively, the required sample size can be calculated using simulation-based approaches.[7] These methods approximate the sampling distribution of $\hat{\theta}_n$ through repeated sampling, model fitting, and validation, and calculate the required sample size by optimising a chosen criterion. In particular, they follow the process outlined in Section 2 and consist of three components:

1) *Specification of the data-generating mechanism*
   - Specify anticipated inputs: model strength, outcome prevalence and the number of predictors.
   - Define $P(\boldsymbol{X};\boldsymbol{\varphi}_2)$ and $P(Y|\boldsymbol{X};\boldsymbol{\varphi}_1)$ and specify parameter values that correspond to the required inputs.
2) *Sampling - training - validation to obtain the sampling distribution of $\hat{\theta}_n$*
   a) Specify the chosen model $M$ and fitting method.
   b) Calculate $\theta_{\text{opt}}$ using very large training and validation datasets.
   c) Generate training and validation datasets, $\mathcal{D}^n$ and $\mathcal{V}^m$, where $m$ is very large.
   d) Fit the model on $\mathcal{D}^n$ and compute $\hat{\theta}_n$ on $\mathcal{V}^m$.
   e) Repeat steps c) - d) $n_{\text{sim}}$ times to obtain the sampling distribution of $\hat{\theta}_n$.
3) *Optimisation*
   Find $n$ to ensure that performance will be close to $\theta_{\text{opt}}$. For example, in terms of the expected CS, $\hat{n} = \arg\min_n |\mathbb{E}(\hat{s}_n) - 0.9|$, although in practice we use $\hat{n} = \arg\min_n |\text{Med}(\hat{s}_n) - 0.9|$ to avoid the effect of outliers.

Simulation-based approaches are flexible and can accommodate any model class, fitting method or performance measure. Also, they can allow consideration of both the average performance and the variability in performance. However, they may require careful specification of the DGM and can be computationally intensive, particularly if the fitting method in Step 2d) above is slow to run.

### 3.3 Assumptions for sample size calculations

The validity of sample size calculations for prediction model development depends on several assumptions described below:

a) *Consistency of modelling approach*
   The model assumed for the sample size calculation should match the model used in development. For example, sample size calculations derived for regression models[5] are not directly applicable to machine learning models (e.g. random forests). The latter may require substantially larger sample sizes to achieve comparable performance [16, 17].

b) *Consistency of fitting method*
   Within a given model class, the fitting method for model development must also be the same as the fitting method assumed for the sample size calculation. For instance, analytical calculations for regression models based on MLE may yield slightly conservative sample sizes if the final model is fitted using penalised methods such as ridge or LASSO[15].

c) *Correct specification of anticipated inputs*

Inputs such as model strength and outcome prevalence must be specified appropriately, ideally using information from prior prediction models and/or analyses. For instance, if the anticipated C-statistic is higher than what can actually be achieved, the sample size will be underestimated, and vice versa. Hence, it is advised that researchers be conservative with these choices.

4) *Absence of distributional drifts*
The population assumed at the sample size calculation stage (i.e. the DGM for $P(Y, \boldsymbol{X}; \boldsymbol{\varphi})$) should match the population used for model development. This requires stability in: i) predictor distribution, $P(\boldsymbol{X}; \boldsymbol{\varphi}_2)$, ii) outcome-predictor relationship, $P(Y|\ \boldsymbol{X}; \boldsymbol{\varphi}_1)$, and iii) marginal outcome distribution, $P(Y)$, e.g. outcome prevalence.

5) *Specification of the data generating mechanism*
Analytical methods only require anticipated summary inputs (e.g. C-statistic, outcome prevalence, number of predictors). Equivalent simulation-based calculations additionally require specification of $P(Y|\boldsymbol{X})$ and $P(\boldsymbol{X})$ that matches the anticipated summary inputs. For aggregate performance measures, results were seen to be robust to misspecification of those distributions, as long as the anticipated inputs are matched[7]. In contrast, more careful specification of $P(Y|\boldsymbol{X})$ and $P(\boldsymbol{X})$ is required for sample size calculations targeting variability of individual predicted probabilities[8].

### 3.4 Simulation settings for this paper - Binary outcomes with logistic regression

For the remainder of this paper we focus on binary outcomes modelled with logistic regression. Our simulation settings are described briefly below, and in more detail in the Supplement.

*Step 1: DGM*

The true outcome model is $\text{logit}\big(P(Y=1|\boldsymbol{X};\boldsymbol{\beta})\big) = \eta = \beta_0 + \boldsymbol{X}^\top \boldsymbol{\beta}_1$, where $X_j \sim N(0,1)$ and $\beta_j = \sqrt{\sigma^2/p}\ , j = 1, \dots, p$. This implies the linear predictor $\eta \sim N(\mu, \sigma^2)$ with parameters chosen to match the desired C-statistic and outcome prevalence.

*Step 2: Sampling distribution of $\hat{\theta}_n$*

We generate development datasets ($\mathcal{D}^n$) and large validation datasets ($\mathcal{V}^m$) with $m = 100{,}000$. The chosen model $M$ for $P(Y|\boldsymbol{X})$ has the same form as the true outcome model. We fit the model to $\mathcal{D}^n$ to obtain $\widehat{\boldsymbol{\beta}}_n$. Unless otherwise stated, the default fitting method is MLE. We then calculate the calibration slope, C-statistic and MAPE using $\mathcal{V}^m$. For example, $\hat{s}_n$, is calculated by fitting the following model to $\mathcal{V}^m$: $\text{logit}\big(P(Y_{\text{val}}=1)\big) = a_{0n} + s_n\, \hat{\eta}_{\text{val}}$, where $\hat{\eta}_{val} = \hat{\beta}_{0n} + \boldsymbol{X}_{\text{val}}^\top \widehat{\boldsymbol{\beta}}_{\boldsymbol{1n}}$. To obtain the sampling distribution of $\hat{\theta}_n$, we repeat this process $n_{\text{sim}} = 2000$ times to ensure a small Monte Carlo simulation error.

*Step 3: Optimisation*

We find the sample size required to achieve a specified target, e.g. $\text{Med}(\hat{s}_n) = 0.9$ or $\Pr(0.85 \leq \hat{s}_n \leq 1.15)$.

In each of the sections that follow, we specify parameter values for each specific scenario. The simulation-based sample size calculations have been implemented in the R package '`samplesizedev`'. For a given sample size this allows us to obtain the sampling distribution of $\hat{\theta}_n$ for a range of performance measures.

# 4 A heuristic bias reduction adjustment to Riley's sample size formula for the expected calibration slope

## 4.1 Explaining bias in Riley's sample size formula for the expected calibration slope

Limiting the expected degree of model overfitting, typically quantified through the expected calibration slope, provides a sensible starting point for sample size calculations in prediction modelling. For a logistic regression model of the form (1), $\text{logit}\big(P(Y|\boldsymbol{X};\boldsymbol{\beta})\big) = \beta_0 + \boldsymbol{X}^\top\boldsymbol{\beta}_1$, Riley et al. (2019)[5] derived the following 'calibration formula'

$$n = \frac{p}{(S-1)\log\left(1 - \frac{R_{CS}^2}{S}\right)} \tag{2}$$

which provides the required sample size for the expected calibration slope $\mathbb{E}(\hat{s}_n)$ to equal a target value $S$, for a model with $p$ predictor parameters and Cox-Snell adjusted $R^2$, $R_{CS}^2$. A commonly used value for $S$ is 0.9, close to the ideal value of 1. The derivation for the calibration formula uses the 'heuristic shrinkage' estimator of van Houwelingen (1990)[18] which provides an approximation for the calibration slope:

$$\hat{s}_n \approx 1 - \frac{p}{\Delta\chi^2}, \tag{3}$$

where $\Delta\chi^2$ is the likelihood ratio test statistic from fitting a logistic regression to a sample of size $n$. This approximation is closely related to the results of Copas (1983, 1997)[19 20].

The calibration formula implies that for a given $p$, stronger models (higher $R_{CS}^2$/C-statistic) require smaller sample sizes to achieve a given calibration target. However, equation (2) systematically underestimates the required sample size when discrimination is high, due to bias in approximation (3)[21]. Empirically, sample sizes calculated using (2) may need to be inflated by 20%, 50% and 100% for $C = 0.8, 0.85$ and $0.9$, respectively[7], in order to meet the target calibration slope.

Our proposed correction that follows is motivated by considering the relationship between logistic regression and linear discriminant analysis (LDA). In modelling the association between $Y$ and $\boldsymbol{X}$ we distinguish between logistic regression, which models $Y|\,\boldsymbol{X}$, and LDA, which models $\boldsymbol{X}|Y$. The first implies a 'prospective' DGM (generate $Y$ given $\boldsymbol{X}$) while the second, a 'retrospective' DGM (generate $\boldsymbol{X}$ given $Y$). When the true DGM is retrospective and the assumptions of LDA are met ($\boldsymbol{X}|Y$ normal and common variance covariance matrix $\Sigma$), then the regression coefficients in model (1) coincide with the log-odds ratios in the corresponding LDA model. In contrast, when the true DGM is prospective and $\boldsymbol{X}$ is normal, approximate equivalence holds only when class separation (model discrimination) is modest. We provide more details about the relationship between LDA and logistic regression under prospective and retrospective DGMs in the Supplement.

*Understanding bias in the heuristic approximation for high discrimination*

A key assumption for the validity of equation (3) is that there is limited variability among the weights $\hat{\pi}_i(1-\hat{\pi}_i), i = 1, \dots, n$, in the Fisher information for $\widehat{\boldsymbol{\beta}}$. This assumption holds when the degree of discrimination in the data is modest. As Copas (1983)[19] noted, for (marginally) normally distributed predictors, this coincides with the conditions under which the log odds ratios, $\widehat{\boldsymbol{\delta}}_{\boldsymbol{1}}$, obtained from fitting an LDA model to data from a prospective DGM are asymptotically equal to the logistic regression coefficients, $\widehat{\boldsymbol{\beta}}_1$. When discrimination is high, the range of the fitted probabilities increases, causing the weights $\hat{\pi}_i(1-\hat{\pi}_i)$ to become more heterogeneous. As a result, approximation (3) overestimates the calibration slope leading to underestimation of the required sample size to achieve a target expected calibration slope. The degree of overestimation in $\hat{s}_n$ is reflected by the attenuation of the LDA coefficients $\widehat{\boldsymbol{\delta}}_1$ relative to $\widehat{\boldsymbol{\beta}}_1$[7]. We exploit this discrepancy to derive a heuristic correction to equation(2) by adjusting the model strength measure used as input, so that it reflects the effective information available in the data.

### 4.2 Proposed Adjustment

As $R^2_{CS}$ is not routinely reported, Riley and Collins (2021)[22] proposed deriving it from the more accessible inputs C-statistic, $C$, and outcome prevalence, $\phi$. Their approach used *a retrospective DGM*, consistent with an LDA model which assumes conditional normality with equal variances. In particular, they considered the linear predictor $\eta$ expressed via the conditional distributions $\eta|Y = 1 \sim N(d, 1)$ and $\eta|Y = 0 \sim N(0,1)$, where $d = \sqrt{2}\,\Phi^{-1}(C)$[23] and $P(Y = 1) = \phi$. Under these assumptions logistic and LDA log-odds ratios coincide asymptotically. So, they obtained $R^2_{CS}$ by fitting a logistic model to a large sample from this DGM. While this approach facilitates calculation of $R^2_{CS}$, it does not address the underestimation bias in (2) for high discrimination. Replacing the original C-statistic, $C$, with a more conservative value, $C_{\text{adj}}$, when discrimination is high would lead to larger sample sizes. We next discuss what suitable values for $C_{\text{adj}}$ might be.

We propose informing the choice of $C_{\text{adj}}$ by following a similar strategy to [22], but instead considering a *prospective DGM* with $\eta \sim N(\mu, \sigma^2)$ and $Y \sim \text{Bernoulli}(\text{logit}^{-1}(\eta))$. The values of $\mu$ and $\sigma^2$ are chosen to match the anticipated $C$ and $\phi$. Then, $\eta$ can be expressed as the linear combination of $p$ independent standard normal variables $X_j$: $\eta = \beta_0 + \boldsymbol{X}^{\top}\boldsymbol{\beta}_1$ where $\beta_{1j} = \sigma/\sqrt{p}$, $j = 1, \dots, p$ and $\beta_0 = \mu$. Under this prospective DGM, the LDA log-odds ratios, $\delta_{1j}$, do not generally coincide with logistic regression coefficients $\beta_{1j}$. We can characterise this discrepancy by generating a large dataset from this prospective DGM and fitting an LDA model to obtain $\hat{\delta}_{1j}$. Unless model discrimination is modest, $|\hat{\delta}_{1j}| < |\hat{\beta}_{1j}|$, with the discrepancy increasing as model discrimination increases. This is because marginal normality of $X_j$ does not imply conditional normality $X_j|Y$. As the degree of attenuation of $|\hat{\delta}_j|$ relative to $|\hat{\beta}_{1j}|$ mirrors the bias in the calibration formula, we use it to inform the choice of $C_{\text{adj}}$. To achieve this, we define an adjusted linear predictor $\eta_{\text{adj}} = \hat{\delta}_0 + \boldsymbol{X}^{\top}\hat{\boldsymbol{\delta}}_1$ and compute the corresponding C-statistic, $C_{\text{adj}}$, based on this. Since $\text{Var}(\eta_{\text{adj}}) < \text{Var}(\eta)$, $C_{\text{adj}} < C$.

We propose replacing the actual (anticipated) value of $C$ with the adjusted value, $C_{\text{adj}}$, which can be computed easily using the full algorithm provided in the Supplement or directly using the `c_adj` function of the package samplesizedev. For example, for $\phi = 0.1$, $C = 0.85, p = 10$, `c_adj(target.prev = 0.1, target.c = 0.85, p = 10)` gives $C_{\text{adj}} = 0.804$.

### 4.3 Evaluation of the proposed adjustment

We used simulation to assess whether replacing $C$ with $C_{\text{adj}}$ reduces bias in the calibration formula. We considered scenarios with $\phi \in \{0.1, 0.3, 0.5\}$, $p \in \{5, 10, 15, 20\}$ and $C \in \{0.7, 0.75, 0.8, 0.85, 0.9\}$. We computed the required sample size using equation (2) targeting an expected calibration slope of $S = 0.90$ using i) the actual C-statistic, $C$ and ii) the adjusted C-statistic, $C_{\text{adj}}$. The corresponding sample sizes are denoted by '$n_{\text{orig}}$' and '$n_{\text{adj}}$', respectively. For each sample size we then calculated $\text{Med}\left(\hat{s}_{n_{\text{orig}}}\right)$ and $\text{Med}\left(\hat{s}_{n_{\text{adj}}}\right)$ and assessed their closeness to the target value of 0.90. The results are presented in Table 2 for $p = 10$. Full results are in Table S1.

[Table 2]

The deviation between $C$ and $C_{\text{adj}}$ increased with increasing $C$ and $\phi$, consistent with the bias in equation (2). Using $C_{\text{adj}}$ as the input improved the median calibration slope substantially for $C \geq 0.8$, the range of $C$ values for which a correction is most needed. Indicatively, when $\phi = 0.1$, for $C = 0.8, 0.85$ and $0.9$, the adjusted C-statistic was $C_{\text{adj}} = 0.770$, 0.804, and 0.834, leading to an inflation of the original sample size by 26%, 42% and 64%, respectively. This inflation led to a median calibration slope substantially closer to 0.9, although for the highest model strength, $C = 0.9$, the sample size was still slightly underestimated. For values of $C < 0.75$ the effect of the correction was less pronounced. The results were similar for other numbers of predictors (Table S1).

## 5 The importance of accounting for variability in performance in sample size calculations

### 5.1 Adapted sample size calculations targeting the Probability of Acceptable Performance

To appreciate the particularities of sample size calculations for developing prediction models, it is instructive to contrast them with conventional sample size calculations. Power- and precision-based calculations assume that the fitting method yields an approximately unbiased estimate of the effect of interest, the *estimand*, and focus on estimating it with sufficient precision. In prediction modelling, the estimand is a predictive performance measure $\theta$, such as the calibration slope, C-statistic, MAPE etc. Importantly, the expected value of an estimator for $\theta$ *depends on the development sample size,* $n$. Consequently, sample size calculations for prediction models often target the expectation of $\hat{\theta}_n$. As $n$ increases, $\mathbb{E}(\hat{\theta}_n) \rightarrow \theta_{\text{opt}}$, while the sampling distribution of $\hat{\theta}_n$ becomes increasingly concentrated around $\theta_{\text{opt}}$. For example, Figure S1 shows the sampling distribution of the calibration slope for different sample sizes at specific values of the C-statistic, number of predictors and prevalence ($C = 0.8, \phi = 0.1, p = 10$). As sample size increases, not only does $\text{Var}(\hat{s}_n) \rightarrow 0$, but also $\mathbb{E}(\hat{s}_n) \rightarrow 1$. This suggests that sample size criteria for prediction model development should ensure both that the model's average performance is sufficiently close to the optimal performance *and* that the variability in performance is acceptably small.

To achieve this, we propose an adaptation of existing sample size calculations that aims to ensure that the probability of acceptable performance, $\text{PrAP}(\hat{\theta}_n) = P(l_\theta \leq \hat{\theta}_n \leq u_\theta)$ is sufficiently high, for a prespecified acceptability range $[l_\theta, u_\theta]$. An analogous approach may be considered for the individual predicted probability $\hat{\pi}_n(\boldsymbol{x}^*)$, for an individual with predictor values $\boldsymbol{x}^*$. We illustrate this issue using the calibration slope as the performance measure. Without loss of generality, we may define the Probability of Acceptable Performance in terms of the CS ($\text{PrAP}(\text{CS}) = \text{PrAP}(\hat{s}_n) = \text{Pr}(0.85 \leq \hat{s}_n \leq 1.15)$ and calculate the required sample size such that this probability meets a threshold, e.g. 80%.

### 5.2 Example – Demonstrating the importance of accounting for the variability in performance

To demonstrate the importance of accounting for variability in the CS we performed a simulation study with the settings described in Section 3.4. The C-statistic and outcome prevalence were fixed at $C = 0.7$ and $\phi = 0.1$, while the number of predictors varied between $p = 4$ and $28$. For each predictor scenario, we calculated the required sample size using the simulation-based approach, using MLE as the fitting method, such that either i) $\text{Med}(\hat{s}_n) = 0.9$ or ii) $\text{PrAP}(\hat{s}_n) = \text{Pr}(0.85 \leq \hat{s}_n \leq 1.15) = 0.8$. We used the R package `samplesizedev`. At the recommended sample sizes for each predictor scenario, we obtained the sampling distributions of the CS, C-statistic and MAPE using the simulation-based approach, summaries of which are presented in Figure 1 and Figure 2.

***Standard calculation.*** Figure 1A shows the CS when the sample size was estimated to achieve a median CS value of 0.9 (Standard calculation). The vertical line-segments correspond to 95% central intervals (2.5th - 97.5th percentile). Even though the target calibration slope of 0.9 is achieved on average, the variability is large when the specified number of predictors is small, due to the relatively small sample size. This suggests that our confidence in obtaining a model with CS close to the target value of 0.9 is lower when a smaller model is specified. To quantify the effect of this increased variability we calculated the proportion of times the CS falls in the prespecified interval [0.85,1.15]. Figure 1B is the plot of the values just calculated. The proportion of times the CS is in [0.85-1.15] is much lower when the specified number of predictors is small. For example, it is close to 73% when $p = 12$, but it is only 54% when $p = 4$. These findings have important practical implications. Since smaller models require smaller sample sizes to achieve a target calibration slope on average (for fixed outcome prevalence and C-statistic), the associated variability can be substantial. As a result, such models may be unstable and unreliable in practice.

***New calculation.*** To address this limitation, we then directly targeted $\mathrm{PrAP}(\hat{s}_n)$ in the sample size calculation. As shown in Figure 1C, calculating the sample size to ensure that $\mathrm{PrAP}(\hat{s}_n) = 0.8$ (blue lines) results in considerably larger sample sizes when the number of predictors is small. For example, as shown in Figure 1D, when $p = 4$, the required sample size to achieve $\mathrm{PrAP}(\hat{s}_n) = 0.8$ is 2.7 times larger than the corresponding calculation that aims at $\mathrm{Med}(\hat{s}_n) = 0.9$. In contrast, for $p = 18$ or more predictors, the required sample sizes were similar under the two approaches.

[Figure 1]

Figure 2 shows summaries of the sampling distribution of the C-statistic and MAPE at the recommended sample sizes targeting $\mathrm{Med}(\hat{s}_n) = 0.9$ and $\mathrm{PrAP}(\hat{s}_n) = 0.8$, respectively. Similar to the results seen for the CS, for sample sizes corresponding to $\mathrm{Med}(\hat{s}_n) = 0.9$, the variability in the C-statistic and MAPE is very high when the number of predictors is small but are controlled much better for sample sizes calculated using $\mathrm{PrAP}(\hat{s}_n) = 0.8$.

[Figure 2]

### 5.3 Choice of interval for acceptable performance and threshold

In practice, the choice of target values for either the average calibration slope or the probability of acceptable performance is subjective. In this illustration, we have shown that aiming at a median calibration slope of 0.9, the default value in current calculations, results in a large variability in the calibration slope for small models. Indicatively, aiming at median calibration slope of 0.9 corresponds to an approximately 80% probability that the calibration slope lies within the interval [0.85-1.15] for a model with 20 predictors (Figure 1). The newly proposed calculation aiming at $\mathrm{PrAP}(\hat{s}_n) = 0.8$ effectively guards against model instability when the specified number of predictors is relatively small (e.g., fewer than 10) by providing a substantially larger sample sizes than the current calculations. For models with 10-30 predictors, which are also common in clinical research, the new calculation results in similar sample sizes as the ones provided by the current calculation for the calibration slope.

The increased variability in the calibration slope is also an indicator of the variability in the individual predicted probabilities, which may ultimately affect clinical decision making. For example, let us consider the situation where $p = 5$, $\phi = 0.2$ and $C = 0.8$. For these input parameters, the required sample size to ensure an expected calibration slope of 0.9 is 300. Figure 3A shows the sampling distribution of the calibration slope for this sample size; this appears to be very wide. Figure 3B shows the median and 95% central interval for the predicted probabilities corresponding to the true probability distribution shown in Figure 3C. If we now consider a specific observation with predictor values $\boldsymbol{x}^*$ that correspond to an individual predicted probability of $\pi(\boldsymbol{x}^*) = 27\%$, the sampling distribution of $\hat{\pi}_{n=300}(\boldsymbol{x}^*)$ is shown in Figure 3D. The distribution is fairly symmetric and wide, indicating that instability in the predicted probabilities due the small development size could potentially affect clinical decisions if, for example, these were based on a pre-specified probability threshold. In this scenario, the sample size could be based on the width of the intervals in Figure 3B, e.g. such that the majority of the intervals are sufficiently narrow. Further discussion on sample size calculations for individual predicted probabilities can be found in Riley et al. (2026).[8]

[Figure 3]

### 5.4 Analytical approach to approximate the sample size for PrAP for the calibration slope

In this section we propose an approximate analytical approach to calculate the sample size for $\mathrm{PrAP}(\hat{s}_n)$ that is computationally efficient. Our approach combines ideas from the closed-form expressions of Riley et al. (2019)[5] and Pavlou et al. (2021)[24] which calculate the sample size aiming at $\mathbb{E}(\hat{s}_n)$ and the variance of the estimated CS in validation data, respectively. We show how we can approximate $\mathrm{Var}(\hat{s}_n)$ based on $n, p, \phi, C$ and $\mathbb{E}(\hat{s}_n)$. We then use this result to obtain $\mathrm{PrAP}(\hat{s}_n)$ for a given sample size and finally show how we can compute the sample size aiming at a target $\mathrm{PrAP}(\hat{s}_n)$ for a given acceptability interval $(l_s, u_s)$.

#### *5.4.1 Closed-form expressions to approximate $Var(\hat{s}_n)$ and $PrAP(\hat{s}_n)$*

With a view to developing an analytical sample size calculation that targets $\mathrm{PrAP}(\hat{s}_n)$ we now discuss analytical expressions to approximate $\mathrm{Var}(\hat{s}_n)$ and subsequently $\mathrm{PrAP}(\hat{s}_n)$ using only information on $n, C, \phi$ and $\mathbb{E}(\hat{s}_n)$. Suppose that a model has been developed using a sample of size $n$ with an expected CS, $\mathbb{E}(\hat{s}_n)$. As discussed in the Supplement, a useful empirical approximation for the variance of the sampling distribution of $\hat{s}_n$, $\mathrm{Var}(\hat{s}_n)$, is

$$\mathrm{Var}_{\mathrm{app}}(\hat{s}_n) = \frac{\mathbb{E}(\hat{s}_n)^2}{2\,\phi\,(1-\phi\,)\,n\,\Phi^{-1}(C)^2} + \frac{2\,\mathbb{E}(\hat{s}_n)^2}{n-2}. \tag{4}$$

It is important to note that $\mathbb{E}(\hat{s}_n)$ and $n$ enter equation (4) *jointly*. That is, for given values of $\phi$ and $C$, $\mathbb{E}(\hat{s}_n)$ is the expected calibration slope corresponding to sample size $n$. For given $\phi$ and $C$, the sample size required to achieve a target expected calibration $\mathbb{E}(\hat{s}_n)$ is *larger* for higher number of predictors according to formula (2). Hence, keeping every other input fixed, $\mathrm{Var}_{\mathrm{app}}(\hat{s}_n)$ is smaller for higher number of predictors, in line with the simulation results of the section 5.2. We note that approximation (4) was seen to be slightly biased for $C \geq 0.75$ [24], for the same reasons the calibration equation (2) is biased for high $C$. Hence, the adjustment to the input value of $C$ discussed in Section 4 is applied here too.

We investigated the performance of equation (4) for combinations of $C = 0.65 - 0.9$, $p = 5, 10, 20$ and $\phi$= 0.1, 0.3, 0.5. For each C-statistic, prevalence and number of predictors scenario, we used the simulation-based approach to calculate the sample size required to achieve $\mathrm{PrAP}(\hat{s}_n) = 0.8$. Then, for each combination of prevalence and number of predictors we calculated $\mathrm{SD}_{\mathrm{sim}} = \sqrt{\mathrm{Var}(\hat{s}_n)}$ and $\mathbb{E}(\hat{s}_n)$ using simulation (function `expected_performance` in `samplesizedev`). Finally, using the values of $C, \phi, n$ and $\mathbb{E}(\hat{s}_n)$ we calculated $\mathrm{SD}_{\mathrm{app}}(\hat{s}_n) = \sqrt{\mathrm{Var}_{\mathrm{app}}(\hat{s}_n)}$ using equation (4). The results presented in Table S3 show very good agreement between $\mathrm{SD}_{\mathrm{sim}}(\hat{s}_n)$ and $\mathrm{SD}_{\mathrm{app}}(\hat{s}_n)$ unless model discrimination is very high (C>0.85).

As seen previously[7], the sampling distribution of $\hat{s}_n$ is approximately normal when the sample size is not too small, and, hence, $\mathrm{PrAP}(\hat{s}_n)$ can be approximated reasonably well by:

$$\mathrm{PrAP}_{\mathrm{app}}(\hat{s}_n) = 1 - \left( \Phi\left( \frac{l_s - \mathbb{E}(\hat{s}_n)}{\sqrt{\mathrm{Var}_{\mathrm{app}}(\hat{s}_n)}} \right) + \Phi\left( \frac{\mathbb{E}(\hat{s}_n) - u_s}{\sqrt{\mathrm{Var}_{\mathrm{app}}(\hat{s}_n)}} \right) \right) \quad (5)$$

For example, Figure S2 shows the true sampling distribution of the calibration slope for $C = 0.7,\ \phi = 0.1$ and $p = 10,$ when the sample size is chosen to ensure that $\mathrm{PrAP}(\hat{s}_n) = 0.8$. For these inputs, $\mathbb{E}(\hat{s}_n) \approx 0.92$, and the approximate sampling distribution of $\hat{s}_n$ assuming normality with mean $\mathbb{E}(\hat{s}_n)$ and variance $\mathrm{Var}_{\mathrm{app}}(\hat{s}_n)$ overlaps substantially with the true distribution. This approximation helps us derive an analytical approach for approximating the sample size that corresponds to a target $\mathrm{PrAP}(\hat{s}_n)$.

### 5.4.2 *Analytical approach to calculate the sample size for $PrAP(\hat{s}_n)$*

We now show how relationships (2), (4) and (5) under their respective assumptions, can be jointly used to calculate the sample size, $n$, so that $\mathrm{PrAP}(\hat{s}_n)$ meets a pre-specified target, $t$. The new calculation requires the conventional inputs of the anticipated prevalence $\phi$, the C-statistic, $C$ and the number of candidate predictor variables, $p$, as well as specification of an interval of acceptable calibration $(l_s, u_s)$ and a threshold for $\mathrm{PrAP}(\hat{s}_n)$. Given these inputs, the sample size, $n$ can be obtained by solving the simple optimisation problem below:

1. Provide a value for $\mathbb{E}(\hat{s}_n)$; $\mathbb{E}(\hat{s}_n) = 0.9$ is a reasonable starting value.
2. Calculate:
    a. $n = \frac{p}{(\mathbb{E}(\hat{s}_n)-1)\log\left(1-\frac{R^2_{CS}}{\mathbb{E}(\hat{s}_n)}\right)}$ ;
    b. $\mathrm{Var}_{\mathrm{app}}(\hat{s}_n) = \frac{\mathbb{E}(\hat{s}_n)^2}{2\,\phi\,(1-\phi)\,n\,\Phi^{-1}(C)^2} + \frac{2\,\mathbb{E}(\hat{s}_n)^2}{n-2}$ ;
    c. $\mathrm{PrAP}_{\mathrm{app}}(\hat{s}_n) = 1 - \left( \Phi\left( \frac{l_s-\mathbb{E}(\hat{s}_n)}{\sqrt{\mathrm{Var}_{\mathrm{app}}(\hat{s}_n)}} \right) + \Phi\left( \frac{\mathbb{E}(\hat{s}_n)-u_s}{\sqrt{\mathrm{Var}_{\mathrm{app}}(\hat{s}_n)}} \right) \right)$.
3. If $|\mathrm{PrAP}(\hat{s}_n) - t)| < 0.05$, the required sample size is $n = \frac{p}{(\mathbb{E}(\hat{s}_n)-1)\log\left(1-\frac{R^2_{CS}}{\mathbb{E}(\hat{s}_n)}\right)}$, otherwise update $\mathbb{E}(\hat{s}_n)$ in step 1 and repeat steps 1) - 3) until convergence.

The optimisation is straightforward and fast (few seconds) in standard software, e.g. with the `optim` command in R (implemented in the package `samplesizedev`). We recommend using the adjusted values of $C$ discussed in Section 4 in steps 2a) and 2b) for input $C \geq 0.75$.

To assess the performance of this analytical approach we estimated the sample size required for $\mathrm{PrAP}(\hat{s}_n) = 0.8$ using the simulation-based and analytical approaches for different input values of outcome prevalence ($\phi = 0.1, 0.3, 0.5$), C-statistic ($C = 0.65 - 0.9$) and number of predictors ($p = 5, 10, 20$). Figure 4 shows the results for $p = 10$. The results for the other predictor values are in Figure S3. The analytical approach provided a very good approximation to the true sample size although it slightly underestimated the sample sizes for $C = 0.9$. If the simulation-based approach is preferred, the analytical approach can still be used to provide starting values close to the solution, making the optimisation path much shorter.

[Figure 4]

## 6 The role of shrinkage in model development

As discussed in previous sections, a key aim of sample size calculations is to ensure a sufficiently small degree of model overfitting, as quantified by the calibration slope. The latter is closely linked to shrinkage, particularly when using internal validation in the absence of external data. The use of pre-shrunk estimators has been discussed in the literature as a way of alleviating model overfitting [18, 19, 25]. A common, and arguably the simplest form of shrinkage, is the linear (or uniform) shrinkage factor approach[3], where the regression coefficients are first estimated using MLE, and then uniformly shrunk by a common factor. The shrinkage factor can be estimated using bootstrapping. The intercept may also need to be adjusted to ensure that the average of the predicted probabilities is equal to the outcome prevalence. In this section we focus on this form of shrinkage. Another category of shrinkage methods includes penalised regression approaches such as Ridge and Lasso[26-28] which we consider in the next section in the context of a real data example.

While in principle shrinkage has the potential to reduce model overfitting and improve prediction, caveats have been raised regarding its use in small data-settings [12]. Specifically, although shrinkage can improve calibration slope *on average*, the variability in the CS is increased due to the uncertainty in the estimation of the shrinkage factor. Consequently, the probability of obtaining a model with acceptable calibration performance is not guaranteed to improve when using shrinkage. This probability is a crucial determinant of whether the sample size requirements can be relaxed when shrinkage is used.

### 6.1 Evaluating the performance of post-estimation linear shrinkage factor correction

In this section we explore the effect of shrinkage on PrAP. We used the same simulation settings as in Section 5. For fixed values of $C = 0.7$ and $\phi = 0.1$ and a range of number of predictors scenarios ($4 - 28$) we calculated the required sample size by simulation so that either $\mathrm{Med}(\hat{s}_n) = 0.9$ or $\mathrm{PrAP}(\hat{s}_n) = 0.8$, assuming that the fitting method is MLE. For each of the recommended sample sizes, we obtain the sampling distribution of $\hat{s}_n$ when the fitting method is i) MLE (aligned with the sample size calculation step) or ii) MLE followed by post-estimation shrinkage with a linear shrinkage factor (estimated from 200 bootstrap datasets). The results are shown in Figure 5. Figure 5A shows the median calibration slope (95% central interval) when linear shrinkage was applied.

Regardless of whether the sample sizes were based on $\mathrm{Med}(\hat{s}_n) = 0.9$ or $\mathrm{PrAP}(\hat{s}_n) = 0.8$, the use of shrinkage resulted in median calibration slope close to 1. However, the variability was higher when $p$ was small, especially for the sizes based on $\mathrm{Med}(\hat{s}_n) = 0.9$.

To investigate whether shrinkage can be beneficial despite the high variability, we calculated the probability of acceptable performance in terms of the calibration slope, for each combination of sample size calculation ($\mathrm{Med}(\hat{s}_n) = 0.9$ or $\mathrm{PrAP}(\hat{s}_n) = 0.8$) and fitting method (MLE or MLE+LSF), which is shown in Figure 5B. When the sample size was chosen based on $\mathrm{PrAP}(\hat{s}_n) = 0.8$, the use of shrinkage (green) resulted in visibly improved probability of acceptable calibration for $p > 6$, compared to MLE alone (blue), with the gains being more prominent when the number of predictors was high. When the sample size was chosen based on $\mathrm{Med}(\hat{s}_n) = 0.9$, shrinkage (grey) improved the probability of acceptable calibration compared to MLE alone (red). However, the positive effect of shrinkage was attenuated by the increased variability when the number of predictors was small. For example, when $p = 6$, the probability of acceptable calibration only increased from 60% for MLE alone to 65% for MLE+LSF. Indicatively, the probability of acceptable calibration was less than 80% when the number of predictors was less than 10.

Overall, these results suggest that shrinkage, despite the added variability, can often improve the probability of obtaining a model with acceptable calibration, compared to using MLE alone. However, when the development sample size is small, the increased variability substantially attenuates the benefits of shrinkage. This results in modest improvement compared to using MLE alone, in terms of the probability of acceptable calibration. Consequently, shrinkage *should not*, in general, be used as a means to reduce sample size requirements. Possible exceptions are perhaps situations where the number of predictors is not too small. For example, when the number of predictors was 10 and the fitting method was MLE, the required sample size for $\mathrm{PrAP}(\hat{s}_n) = 0.8$ was 2460. If MLE followed by linear shrinkage were used for model fitting, the target PrAP(CS) would be achieved with a sample size of around 2000.

[Figure 5]

## 7 Case Study – Heart valve surgery

### 7.1 Description

We now demonstrate an application of the sample size calculations proposed in this paper considering data from patients undergoing heart valve surgery[29]. We have used the version of the data studied in [15]. The dataset consists of 16679 individuals in Great Britain and Ireland who had heart valve surgery between 1995 and 2003. The outcome of interest is in-hospital death (binary outcome) following heart valve surgery (prevalence 7%). The data consists of a mixture of eleven binary and continuous variables which were described in detail elsewhere[15]. The aim of this illustration is to demonstrate the application of the proposed methods for calculating the sample size to develop a risk model to predict the risk of in-hospital death.

As the dataset is relatively large, we used sampling without replacement from the data to assess the performance of sample size calculations with the strategy outlined below. The chosen model was logistic regression with the 11 available predictor variables (linear terms and main effects only). The model was fitted to the entire dataset. In bootstrap validation (200 samples) the estimated calibration slope was $\hat{s}_{\text{boot}} = 0.985$ and the optimism-adjusted C-statistic $\hat{C}_{\text{boot}} = 0.73$. As the degree of overfitting was very small, the performance loss compared to the optimal value of $C$, $C_{\text{opt}}$ will be minimal. Hence, $C = 0.73$ is taken to be the optimal value for the model above, and together with the prevalence and the number of predictors is used as input for the sample size calculations.

## 7.2 Sample size calculation

We considered two sample sizes based on:

1) the 'standard' calculation aiming at $\text{Med}(\hat{s}_n)$=0.9: $n_{\text{standard}}$
2) the 'new' calculation aiming at $\text{PrAP}(\hat{s}_n) = \Pr(0.85 \leq \hat{s}_n \leq 1.15) = 0.8$): $n_{\text{new}}$.

The sample sizes were calculated using `samplesizedev` with input values $C = 0.73$, prevalence=0.07 and $p = 11$ predictor using 2000 simulations. This gave $n_{\text{standard}} = 2250$ and $n_{\text{new}} = 2740$.

## 7.3 Resampling process, fitting methods and calculation of performance measures

### *7.3.1 Resampling process*

We used the following process for sampling development and validation datasets to mimic the process that would lead to $\text{Med}(\hat{s}_n)$ and $\text{PrAP}(\hat{s}_n)$. First, we sampled observations without replacement from the original dataset to form a training dataset with the desired sample size ($n_{\text{standard}}$ or $n_{\text{new}}$ ); the remaining data formed a validation dataset. The process was repeated 200 times. The model is fitted on each of the training datasets using the methods described below and validated on the left-out part by calculating measures of predictive performance also described below. The validation datasets were large enough so that the performance measures for a model derived from a given training sample were estimated with small variability.

### *7.3.2 Fitting methods*

The default fitting method was MLE. We first explored whether the sample size calculations above were adequate. To explore whether shrinkage can be helpful in improving predictive performance for the two chosen sample sizes we have considered MLE followed by post-estimation shrinkage with linear shrinkage factor estimated with 200 bootstrap samples ('MLE+LSF') and also ridge regression. For ridge, we used two implementations. First, we used the implementation in `glmnet` with the default tuning method and 10-fold cross-validation for the selection of the tuning parameter ('Standard Ridge'). As standard ridge was previously found to underfit the model, we have also used the modification of Pavlou et al. (2024)[15] in choosing the tuning parameter which can reduce underfitting ('Modified Ridge').

### *7.3.3 Performance measures*

For each sample size and each iteration, we fitted the model with each of the methods on the training dataset and calculated the CS, the C-statistic and the Brier score in the validation dataset. We then summarised the results using boxplots in Figure 6.

## 7.4 Results

The sample size calculation performed as expected, with the median CS and the probability of acceptable calibration close to 0.9 and 0.8, respectively (Figure 6A and B). On average, all shrinkage methods resulted in CS closer to 1 compared to MLE, but with higher variability. They all improved the $\mathrm{PrAP}(\hat{s}_n)$ which is consistent with the simulation results in Section 5.4. The modified tuning approach performed better in terms of CS than Standard Ridge, achieving CS closer to 1 on average and with lower variability (in line with Figure 5B, as the number of predictors (11) is not too small). For the other performance measures considered, the C-statistic and the Brier score, the penalised methods also resulted in somewhat improved performance, although the improvement was less pronounced than for CS.

[Figure 6]

We note that the sample size calculations with the simulation-based approach assumed independent and normal $\boldsymbol{X}'s$, although this assumption can be relaxed by incorporating a distribution for $\boldsymbol{X}$ that resembles the features of the actual data. Further details on this possibility are provided in the Discussion. Nevertheless, as shown in the simulations of Pavlou et al. (2024)[7], provided that the C-statistic and outcome prevalence values are appropriately specified, the distribution of $\boldsymbol{X}$ does not affect the sample size calculation based on the calibration slope, a finding that was also observed in this example.

# 8 Discussion

Sample size calculations for the development of risk models have mainly focused on average performance over repeatedly sampled training samples. The limitation of this approach is that the variability in performance can be high, leading to model instability, and poor performance in individual datasets. Consequently, to ensure that the probability of obtaining acceptable performance in individual development datasets is high, the variability in performance needs to be accounted for. While for a given sample size models with a small number of parameters have been considered to be more stable than larger models[30], our findings show that in the context of sample size calculations it is particularly important to account for variability in performance when the number of predictors in the model is relatively small (e.g. <10).

These considerations have led us to propose an adapted approach to sample size calculation that explicitly accounts for variability in performance by focusing on the probability of acceptable performance in terms of a given measure, e.g. calibration slope, MAPE, C-statistic etc. The implementation of the method is straightforward in the simulation-based framework introduced by Pavlou et al. (2024)[7] via the R package `samplesizedev`. Simulation-based approaches are attractive because they can be tailored to any performance measure and can provide unbiased estimation of the sample size compared to existing analytical formulae which have been found to be severely biased when the predictive strength of the model is high[7]. For a given sample size, `samplesizedev` can provide the entire sampling distribution for a variety of performance measures. Also, given a target for a performance measure, it can calculate the sample size required to meet that target using numerical optimisation. For example, for the calibration slope, this optimisation typically takes 1-3 minutes (2021 MacBook Pro with M2-Pro processor). Although we have primarily focused on aggregate performance, sample size calculations can also target individual predicted probabilities for individuals with specific covariate values[8]. For instance, if a particular probability threshold is intended to be used for clinical decision-making, we may need to ensure that the model can estimate nearby probabilities with minimal bias and variability.

Despite the efficient implementation in our package, simulation-based calculations remain more computationally demanding and slower than methods based on analytical formulae, particularly if one uses them in simulation exercises. However, obtaining analytical expressions for the expectation and variance of a given performance measure is not always straightforward or feasible and usually requires additional assumptions. In this paper we have used an approximate analytical formula for the variance of the calibration slope when the outcome is binary. We then showed how this formula, in combination with the sample size formula for the average calibration slope[5, 18] can be used to obtain the sample size required to achieve a desired probability of achieving acceptable calibration slope. In addition, it can be used to speed up simulation-based sample size calculations. Future work may focus on obtaining approximate expressions for the expectation and variance for other performance measures such as the C-statistic and MAPE.

In this paper, we assumed normality for the distribution of the predictor variables which is unlikely to hold in practice. In practice, simulation-based sample size calculations can be performed with an arbitrary user-defined distribution for $\boldsymbol{X}$, for example the empirical distribution of $\boldsymbol{X}$ to reflect the types of predictors and the relationships between them. Apart from the distribution of $\boldsymbol{X}$, the regression coefficients for the assumed true model would also need to be specified such that anticipated $C$ and prevalence values are met. For example, if an existing dataset is available, an initial model fit can provide an indication of the relative strength of the regression coefficients. Then, the coefficients can be adjusted accordingly (the intercept shifted, and the predictor effects scaled by a common factor) to form a model that matches the anticipated prevalence and C-statistic of the assumed true model. This approach is also possible in the package `samplesizedev`. Nevertheless, Pavlou et al. (2024)[7] investigated various scenarios with different types of predictors and correlations between them, and found that when the summary model characteristics such as prevalence, C-statistic and number of predictors were correctly specified to match that of the assumed true model, the distribution of the predictors had negligible impact on the resulting sample sizes.

Finally, when adhering to the new sample size recommendations, our simulations showed that applying linear shrinkage estimated using bootstrapping resulted in higher probability of acceptable calibration, unless the number of predictors was very small (<6). In our real data application, all shrinkage methods considered were found to improve calibration compared to MLE; other aspects of predictive performance were also improved, albeit to a lesser extent. Based on these results and previous investigations we recommend a conservative approach where the sample size is estimated assuming MLE. However, model-fitting should incorporate some form of shrinkage such as the linear shrinkage factor or modified ridge/lasso, unless the number of predictors is very small.

**Figures and Tables (in the order they appear in paper)**

*Table 1. Notation for the definitions in Section 2*

| Key Notation | Simplified | Explanation |
|---|---|---|
| $Y$ | | Outcome variable |
| $\boldsymbol{X}$ | | Vector of $p$ predictors $\left(X_1, \dots X_p\right)^\top$ |
| $\Pr(Y, \boldsymbol{X}; \boldsymbol{\varphi})$ | $\Pr(Y, \boldsymbol{X})$ | The joint distribution of the outcome and predictors |
| $\Pr(\boldsymbol{X}; \boldsymbol{\varphi_2})$ | $\Pr(\boldsymbol{X})$ | The distribution of predictors |
| $\Pr(Y\|\boldsymbol{X}; \boldsymbol{\varphi_1})$ | $\Pr(Y\|\boldsymbol{X})$ | The distribution of the outcome conditional on predictors |
| $\mathcal{D}^n$ | | Training dataset of size $n$, a sample from $\Pr(Y, \boldsymbol{X}; \boldsymbol{\varphi})$ |
| $\mathcal{V}^m$ | | Validation dataset of size $m$, a sample from $\Pr(Y, \boldsymbol{X}; \boldsymbol{\varphi})$ |
| $\pi(\boldsymbol{x}^*)$ | | Predicted probability for an observation with $\boldsymbol{X} = \boldsymbol{x}^*$ under the true model |
| $M$ | | The chosen type of model for $Y\|\boldsymbol{X}$, e.g. regression, Random Forest, etc. |
| $\theta$ | | Performance measure (e.g. calibration slope, C-statistic) |
| $\boldsymbol{\beta}$ | | Parameters in model $M$ for $Y\|\boldsymbol{X}$, e.g. $\text{logit}\big(\Pr(Y\|\boldsymbol{X})\big) = \beta_0 + \boldsymbol{X}^\top \boldsymbol{\beta}_1$ |
| $\widehat{\boldsymbol{\beta}}_n$ | | Estimated parameters from fitting chosen model on $\mathcal{D}^n$ |
| $\hat{\pi}_n(\boldsymbol{x}^*)$ | | Prediction for an observation $\boldsymbol{x}^*$ from model fitted on $\mathcal{D}^n$ |
| $\widehat{\boldsymbol{\pi}}_n(\mathcal{V}^m)$ | $\widehat{\boldsymbol{\pi}}_n$ | Predictions in the dataset $\mathcal{V}^m$ from model fitted on $\mathcal{D}^n$ |
| $\theta$ | | Performance measure (e.g. calibration slope, C-statistic) |
| $\hat{\theta}(\boldsymbol{\pi}_n, \mathcal{V}^m)$ | $\hat{\theta}_{n,m}$ | Estimated value of $\theta$ from model fitted on $\mathcal{D}^n$ and validated on $\mathcal{V}^m$ |
| $\lim\limits_{m\to\infty} \hat{\theta}_{n,m}$ | $\hat{\theta}_n$ | True value of $\theta$ for model fitted on $\mathcal{D}^n$ and validated on a very large dataset |
| $\lim\limits_{n\to\infty} \hat{\theta}_n$ | $\theta_{\text{opt}}$ | Optimal value of $\theta$ for the chosen model, simplified to $\theta_{\text{opt}}$ |

*Figure 1. **A.** The median calibration slope (CS) (with $2.5^{th}$ and $97.5^{th}$ percentiles) at the sample size required to achieve a median CS of 0.9 (Standard calculation - red). The numbers at the top are these sample sizes. **B.** The probability that the CS lies between 0.85 and 1.15 at the sample sizes based on the Standard calculation **C**. The median CS (with $2.5^{th}$ and $97.5^{th}$ percentiles) at the sample size required to achieve a probability of the CS being in the interval 0.85-1.15 of 80% (New calculation - blue). The numbers at the bottom are these sample sizes. **D**. Required sample size under the Standard and New calculations. Numbers at the top correspond to the ratio of sample sizes for the New versus the Standard calculation. For all predictor scenarios (range: 4-28), the sample size was calculated assuming MLE with the C-statistic and outcome prevalence fixed at 0.7 and 0.1, respectively.*

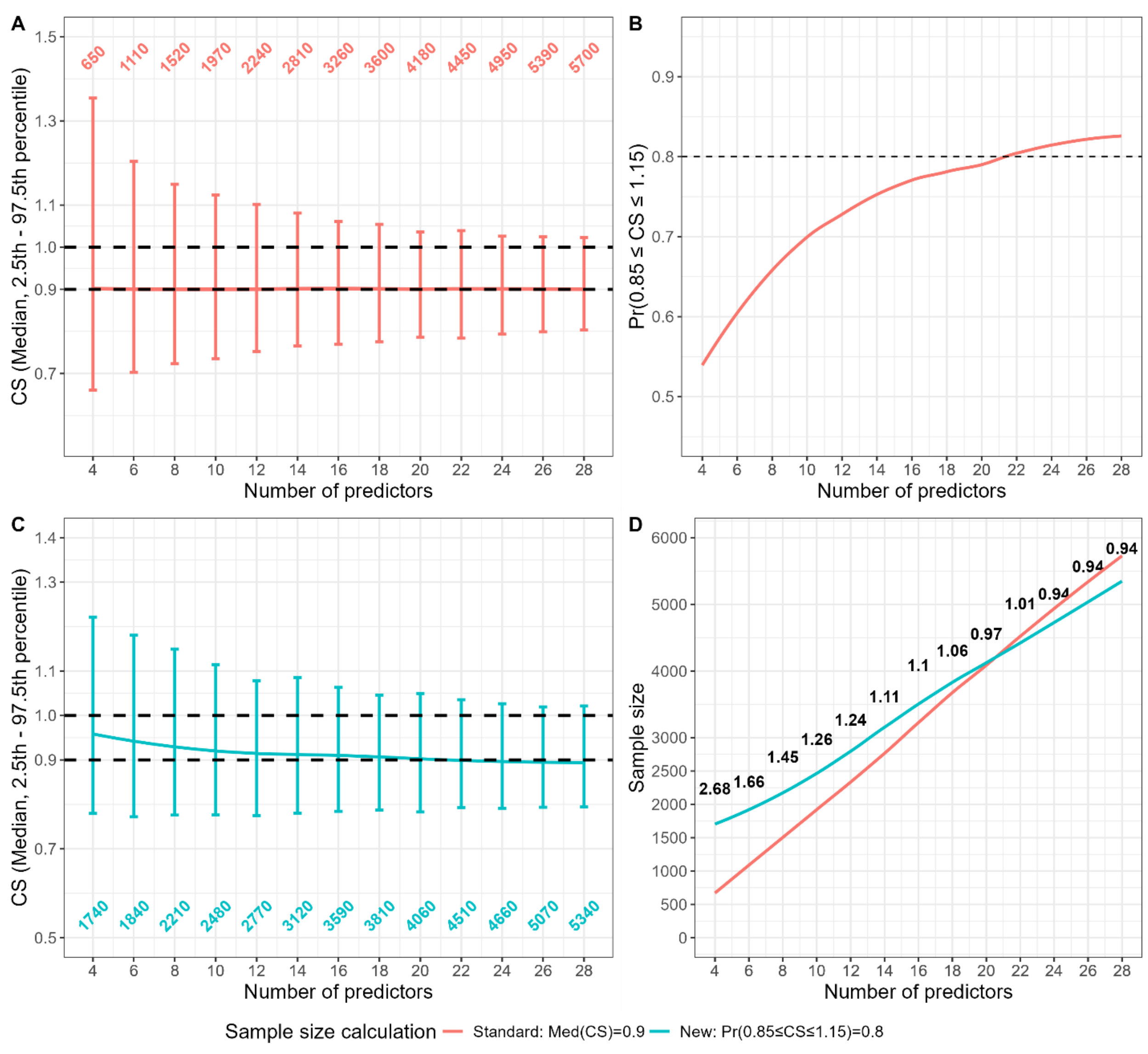

*Figure 2. **A**. Median C-statistic (2.5$^{th}$-97.5$^{th}$ percentiles) at the sample size required i) to achieve a median CS of* 0.9 *(Standard calculation - red) or ii) to achieve a probability of the CS being in the interval 0.85-1.15 of 80% (New calculation - blue). **B**. Median MAPE (2.5$^{th}$-97.5$^{th}$ percentiles) at the sample size based on the Standard or New calculation. For all predictor scenarios (range: 4-28), the sample size was calculated assuming MLE with the C-statistic and outcome prevalence fixed at 0.7 and 0.1, respectively.*

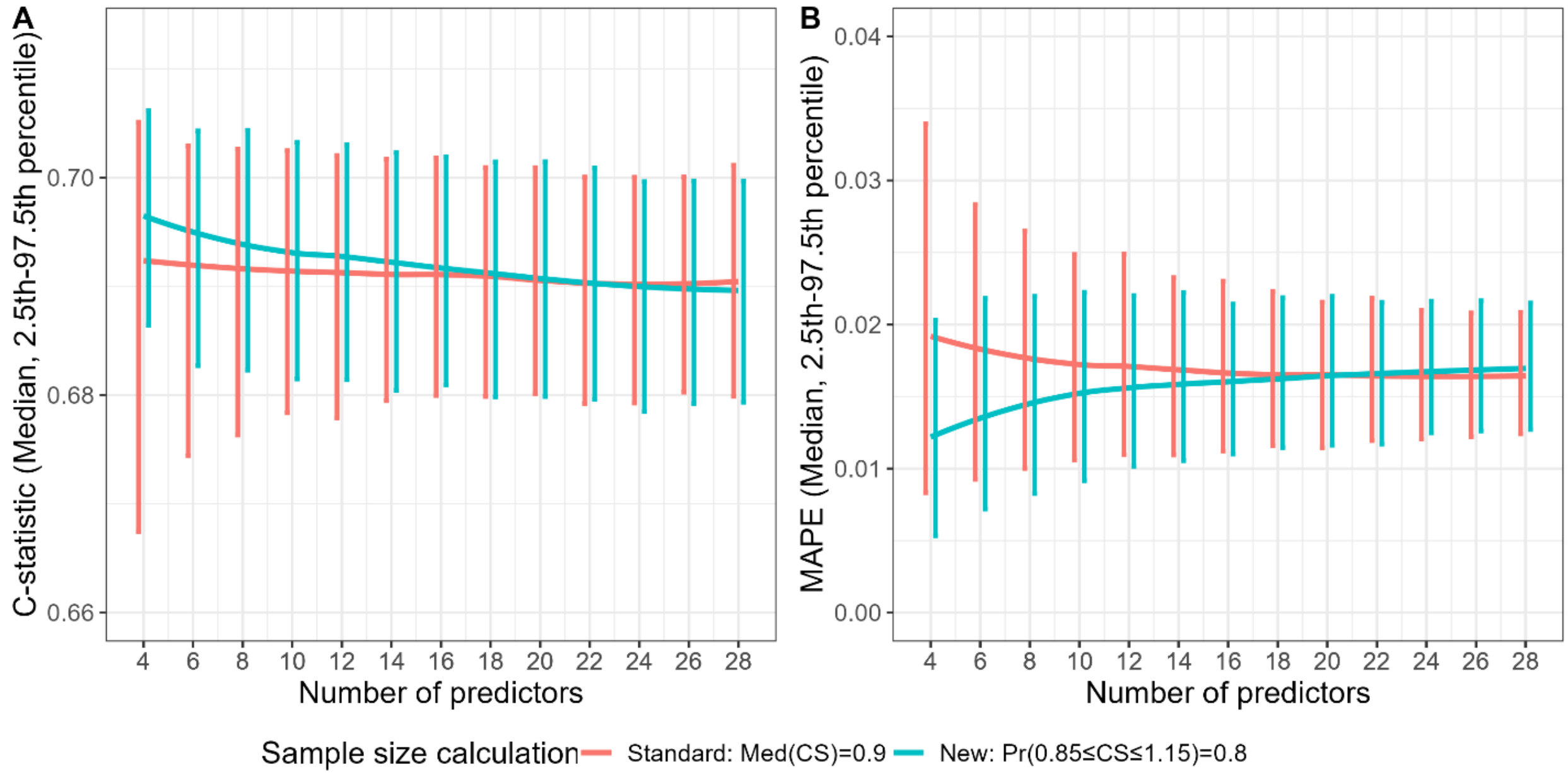

*Table 2. Sample sizes calculated using the analytical calibration formula (MLE is the fitting method) targeting median calibration slope of 0.9, using the i) actual C with corresponding size $n_{orig}$ and ii) adjusted C with corresponding size $n_{adj}$. $Med(\hat{s}_{n_{orig}})$ and $Med(\hat{s}_{n_{adj}})$ denote the median calibration slopes corresponding to sample sizes $n_{orig}$ and $n_{adj}$, respectively. Number of predictors =10.*

| $\phi$ | $C$ | $C_{\text{adj}}$ | $n_{\text{orig}}$ | $n_{\text{adj}}$ | $\text{Med}(\hat{s}_{n_{\text{orig}}})$ | $\text{Med}(\hat{s}_{n_{\text{adj}}})$ | $\frac{n_{\text{adj}}}{n_{\text{orig}}}$ |
|---|---|---|---|---|---|---|---|
| 0.1 | 0.70 | 0.691 | 1,880 | 2,070 | 0.895 | 0.906 | 1.10 |
| 0.1 | 0.75 | 0.733 | 1,160 | 1,350 | 0.890 | 0.906 | 1.16 |
| 0.1 | 0.80 | 0.771 | 760 | 960 | 0.874 | 0.900 | 1.26 |
| 0.1 | 0.85 | 0.804 | 520 | 740 | 0.851 | 0.893 | 1.42 |
| 0.1 | 0.90 | 0.834 | 360 | 590 | 0.803 | 0.881 | 1.64 |
| 0.3 | 0.70 | 0.683 | 820 | 990 | 0.892 | 0.905 | 1.21 |
| 0.3 | 0.75 | 0.720 | 510 | 670 | 0.882 | 0.906 | 1.31 |
| 0.3 | 0.80 | 0.752 | 340 | 500 | 0.858 | 0.899 | 1.47 |
| 0.3 | 0.85 | 0.780 | 240 | 400 | 0.827 | 0.891 | 1.67 |
| 0.3 | 0.90 | 0.805 | 170 | 330 | 0.773 | 0.879 | 1.94 |
| 0.5 | 0.70 | 0.682 | 700 | 850 | 0.883 | 0.905 | 1.21 |
| 0.5 | 0.75 | 0.717 | 440 | 590 | 0.868 | 0.908 | 1.34 |
| 0.5 | 0.80 | 0.748 | 290 | 440 | 0.855 | 0.897 | 1.52 |
| 0.5 | 0.85 | 0.775 | 210 | 360 | 0.815 | 0.892 | 1.71 |
| 0.5 | 0.90 | 0.798 | 150 | 300 | 0.756 | 0.878 | 2.00 |

*Figure 3. A. Sampling distribution of the calibration slope at the sample size required to achieve median CS=0.9. B. Median (2.5th – 97.5th percentiles) for individual predicted probabilities (IPPs) corresponding to a range of true probabilities. The red dashed line indicates the true individual probabilities. C. Distribution of the true probabilities derived from a large dataset from the DGM. D. Sampling distribution for the IPP=0.27 (blue dashed vertical line in C).*

**Sampling Distribution of the CS and Individual Predicted Probabilities (IPPs)**
**Method=MLE, N=300, Prevalence=0.2, C-stat=0.8, No Predictors=5**

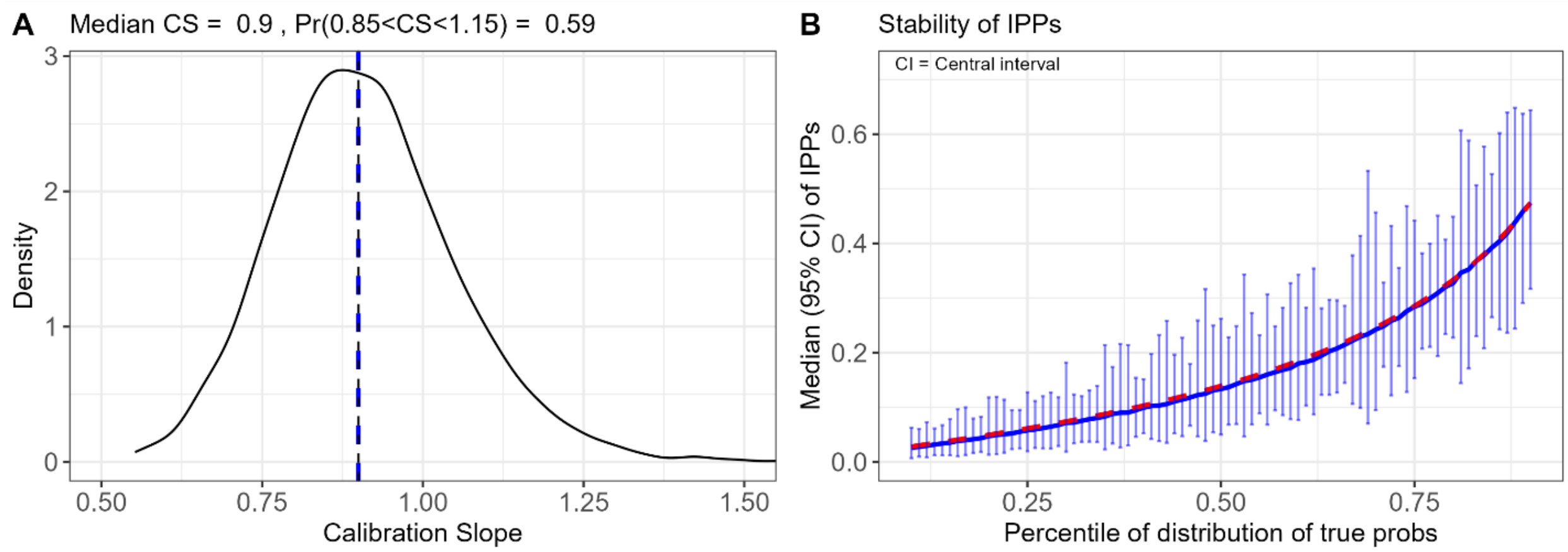

**Probability Distibution of true probabilities and Stability for a specific IPP**

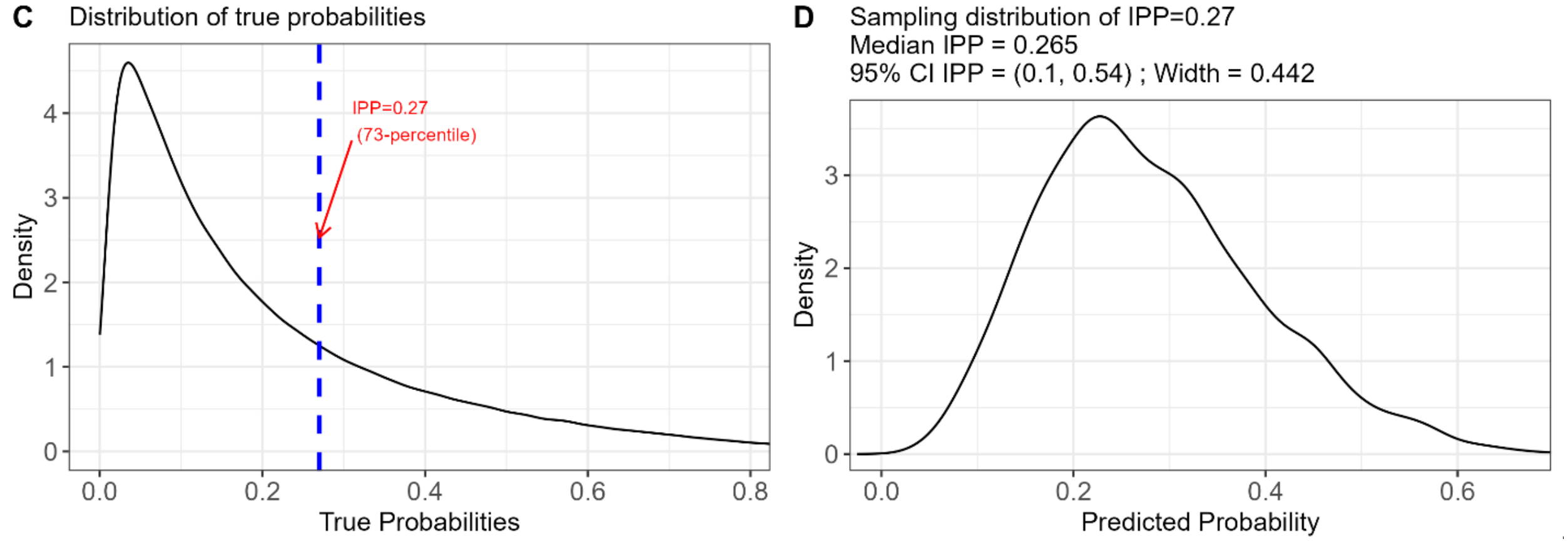

*Figure 4. Sample size required to ensure $PrAP(\hat{s}_n) = 0.8$ using either i) the simulation-based approach (blue) or ii) the analytical approach (purple) for varying values of the C-statistic. For all combinations of the C-statistic (range 0.65-0.9) and outcome prevalence (0.1, 03, 0.5) the sample size was calculated assuming MLE with the number of predictors fixed at 10. Adjusted C-values were used for steps a) and c) of the algorithm for the analytical calculation for $C \geq 0.75$.*

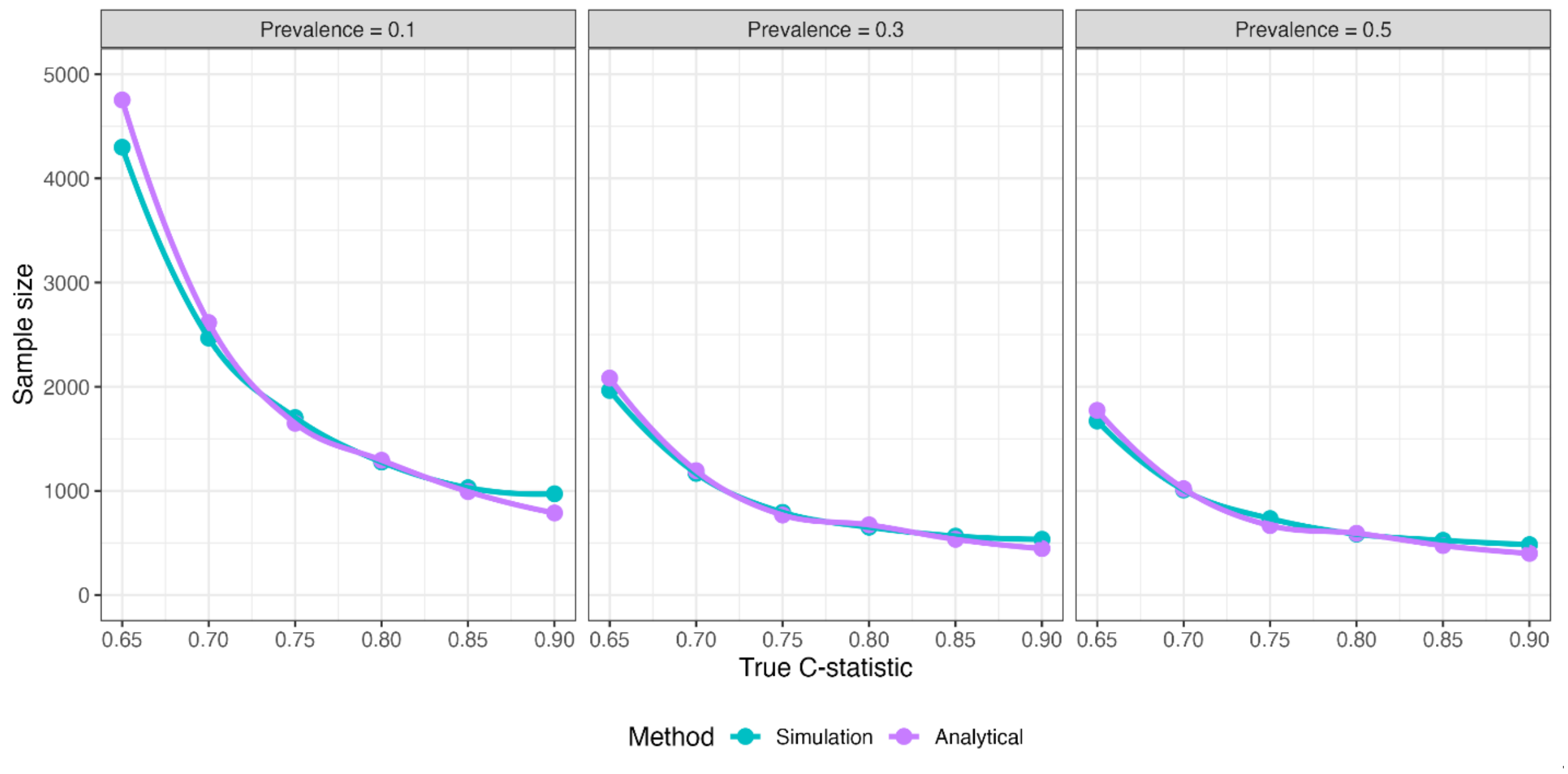

*Figure 5.* ***A.*** *Median Calibration Slope (CS) ($2.5^{th}$ - $97.5^{th}$ percentiles) when linear shrinkage factor (LSF) was used to obtain predictions i) at the sample size required to achieve a median CS of* 0.9 *(Standard sample size calculation + LSF fitting method - grey) or ii) the sample size required to achieve a probability of the CS being in the interval 0.85-1.15 of 80% (New sample size calculation + LSF fitting - green).* ***B.*** *The probability that the CS lies between 0.85 and 1.15 at the sample sizes based on the Standard and New Calculation when the fitting method was either i) MLE (Standard: red and New: blue) or MLE followed by linear shrinkage factor (Standard - grey and New - green). For all predictor scenarios (range: 4-28), the sample size was calculated assuming MLE with the C-statistic and outcome prevalence fixed at 0.7 and 0.1, respectively.*

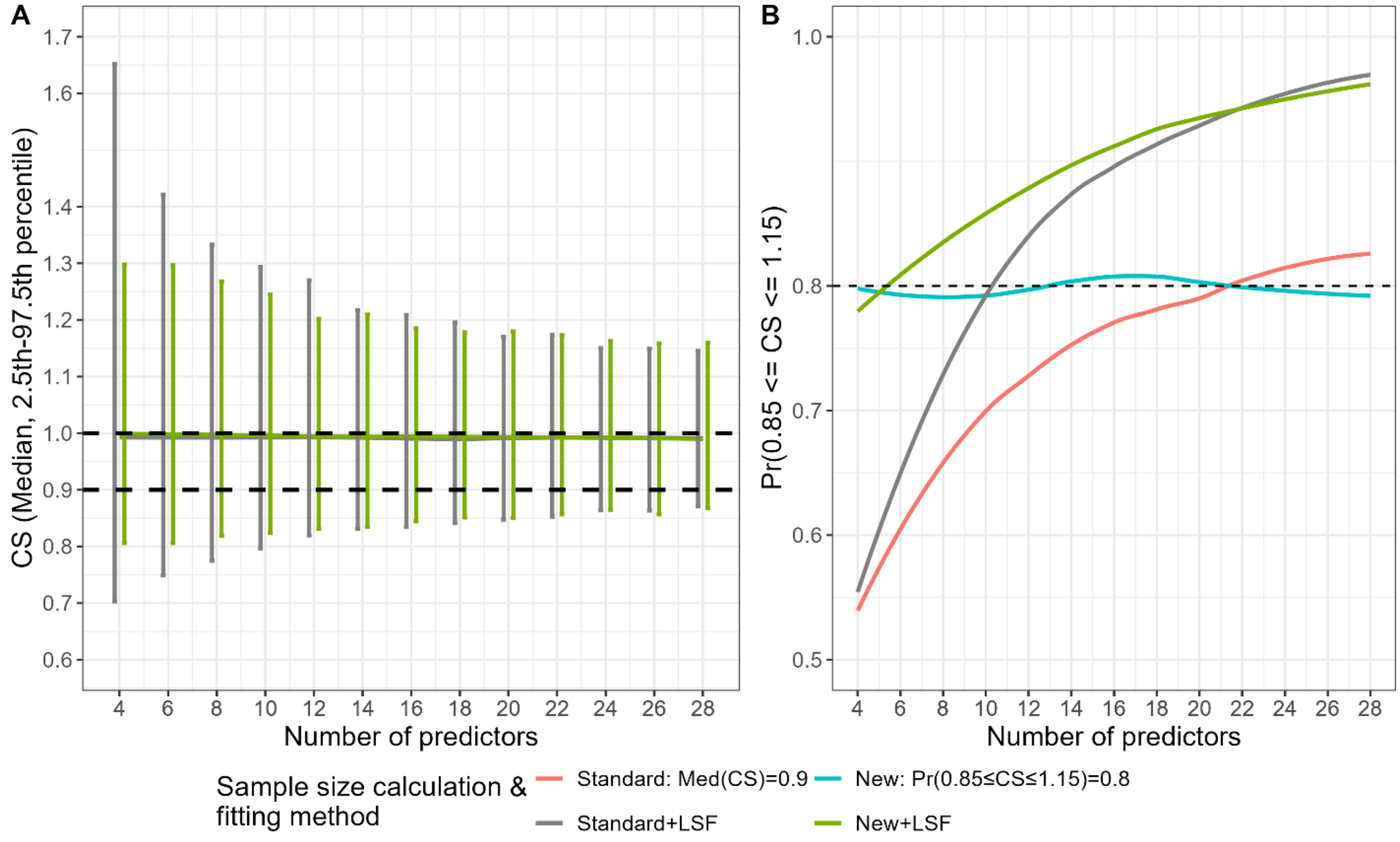

*Figure 6. Measures of predictive performance for the heart valve data. The two sample sizes considered correpond to the Standard and New calculation.*

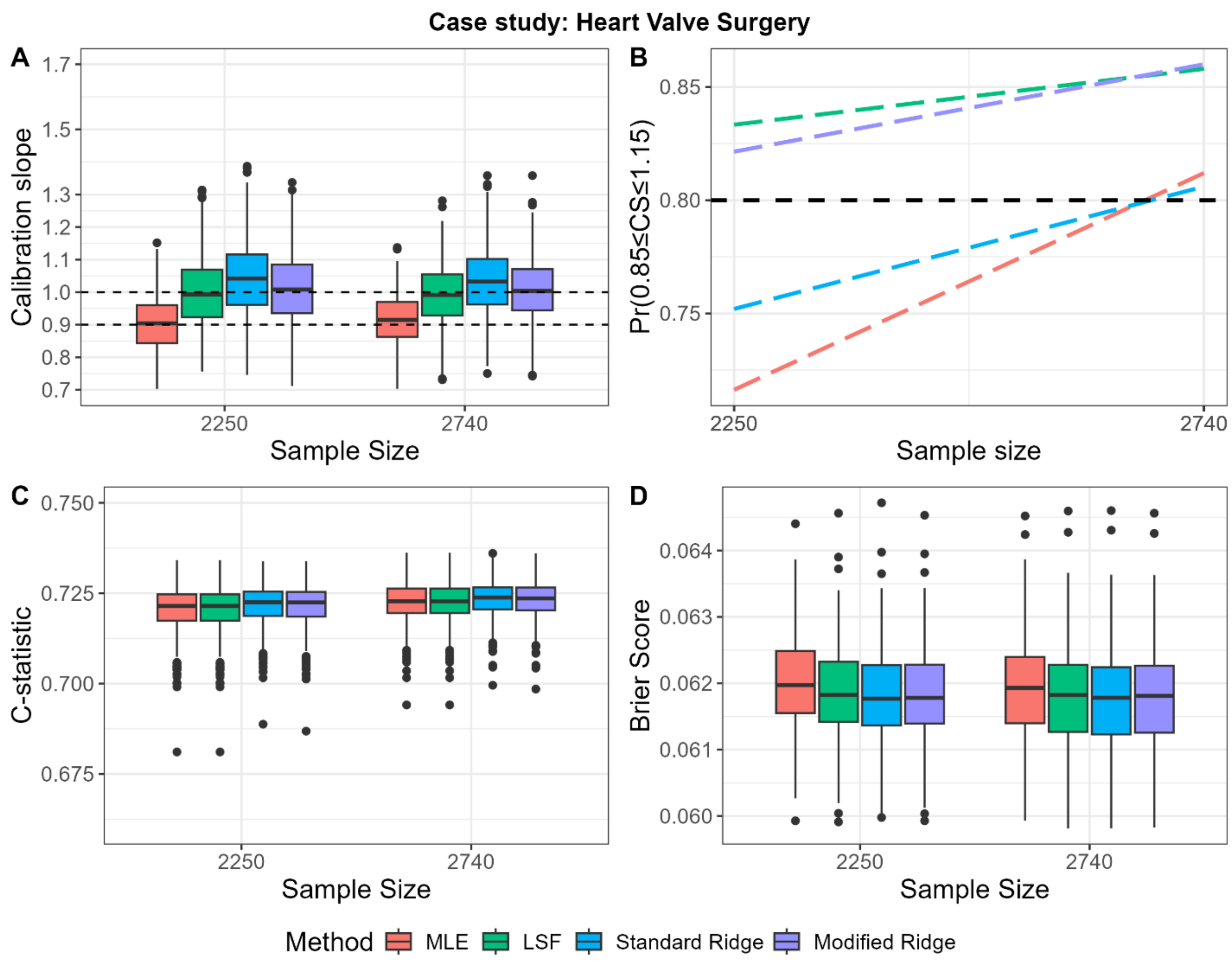

***Supplementary material for the paper***

'Formalising Sample Size Calculations for the Development of Risk Prediction Models: The Importance of Accounting for Performance Variability'

## 1 Supplementary Tables and Figures

Figure S1. ***A.*** *Median Calibration Slope.* ***B.*** *Sampling distribution of the calibration slope. For all sample sizes considered, the C-statistic, outcome prevalence and number of predictors were fixed at 0.8, 0.2 and 10, respectively. The fitting method was MLE.*

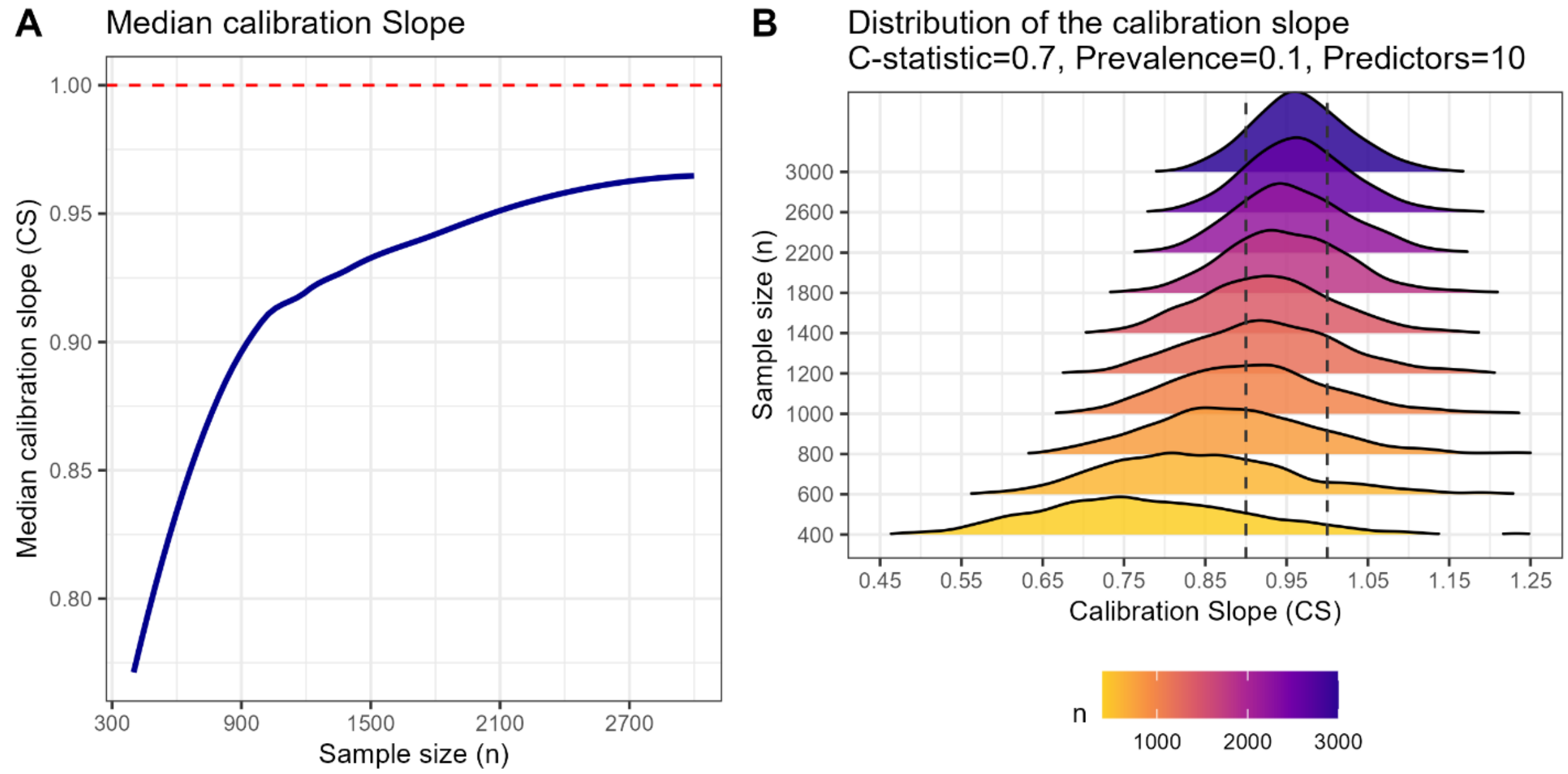

*Figure S2. Distribution of the true and approximate distribution of the calibration slope $\hat{s}_n$. Outcome prevalence = 0.1, C-statistic=0.7, number of predictors = 10.*

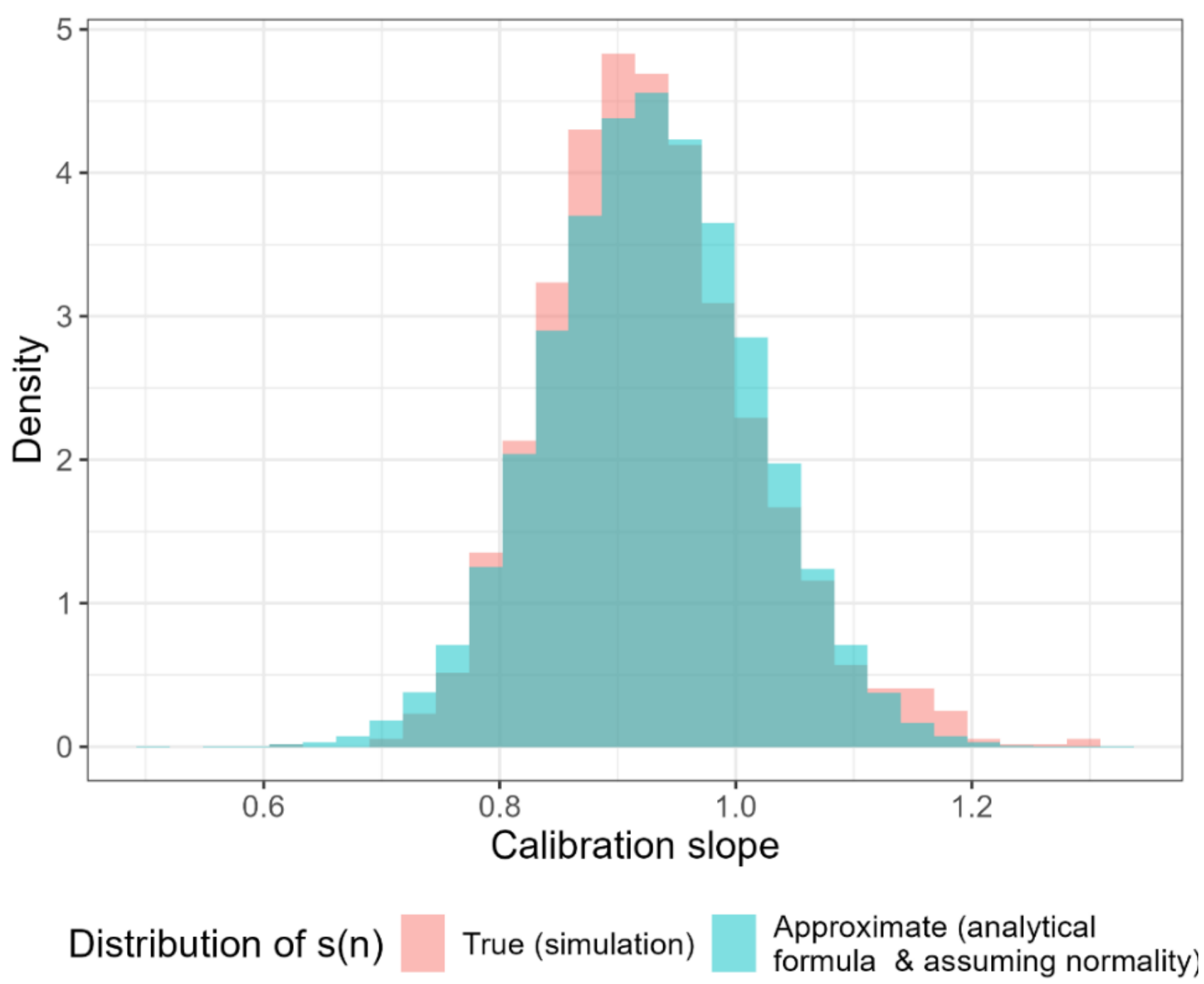

*Table S1. Sample sizes calculated using the analytical calibration formula (MLE is the fitting method) targeting expected calibration slope of 0.9, using the i) actual C* $with\ corresponding\ size\ n_{orig}$ *and ii) adjusted C -*$with\ corresponding\ size\ n_{adj}$. $Med(\hat{s}_{n_{orig}})$ *and* $Med(\hat{s}_{n_{adj}})$ *denote the median calibration slopes corresponding to sample sizes* $n_{orig}$ $n_{adj}$*, respectively.*

| $\phi$ | $p$ | $C$ | $C_{\text{adj}}$ | $n_{\text{orig}}$ | $n_{\text{adj}}$ | $Med(s_{n_{\text{orig}}})$ | $Med(s_{n_{\text{adj}}})$ | $\frac{n_{\text{adj}}}{n_{\text{orig}}}$ |
|---|---|---|---|---|---|---|---|---|
| 0.1 | 5 | 0.70 | 0.692 | 940 | 1,030 | 0.903 | 0.913 | 1.10 |
| 0.1 | 5 | 0.75 | 0.734 | 580 | 670 | 0.893 | 0.914 | 1.16 |
| 0.1 | 5 | 0.80 | 0.772 | 380 | 480 | 0.873 | 0.899 | 1.26 |
| 0.1 | 5 | 0.85 | 0.807 | 260 | 360 | 0.859 | 0.895 | 1.38 |
| 0.1 | 5 | 0.90 | 0.838 | 180 | 290 | 0.809 | 0.880 | 1.61 |
| 0.1 | 10 | 0.70 | 0.691 | 1,880 | 2,070 | 0.895 | 0.906 | 1.10 |
| 0.1 | 10 | 0.75 | 0.733 | 1,160 | 1,350 | 0.890 | 0.906 | 1.16 |
| 0.1 | 10 | 0.80 | 0.771 | 760 | 960 | 0.874 | 0.900 | 1.26 |
| 0.1 | 10 | 0.85 | 0.804 | 520 | 740 | 0.851 | 0.893 | 1.42 |
| 0.1 | 10 | 0.90 | 0.834 | 360 | 590 | 0.803 | 0.881 | 1.64 |
| 0.1 | 15 | 0.70 | 0.692 | 2,820 | 3,090 | 0.901 | 0.903 | 1.10 |
| 0.1 | 15 | 0.75 | 0.733 | 1,730 | 2,030 | 0.888 | 0.901 | 1.17 |
| 0.1 | 15 | 0.80 | 0.770 | 1,140 | 1,460 | 0.871 | 0.897 | 1.28 |
| 0.1 | 15 | 0.85 | 0.803 | 780 | 1,110 | 0.852 | 0.890 | 1.42 |
| 0.1 | 15 | 0.90 | 0.833 | 550 | 890 | 0.806 | 0.878 | 1.62 |
| 0.1 | 20 | 0.70 | 0.692 | 3,760 | 4,120 | 0.893 | 0.904 | 1.10 |
| 0.1 | 20 | 0.75 | 0.732 | 2,310 | 2,740 | 0.883 | 0.901 | 1.19 |
| 0.1 | 20 | 0.80 | 0.769 | 1,520 | 1,960 | 0.870 | 0.898 | 1.29 |
| 0.1 | 20 | 0.85 | 0.803 | 1,040 | 1,490 | 0.852 | 0.887 | 1.43 |
| 0.1 | 20 | 0.90 | 0.832 | 730 | 1,190 | 0.804 | 0.874 | 1.63 |
| 0.3 | 5 | 0.70 | 0.685 | 410 | 490 | 0.893 | 0.913 | 1.20 |
| 0.3 | 5 | 0.75 | 0.723 | 260 | 330 | 0.884 | 0.910 | 1.27 |
| 0.3 | 5 | 0.80 | 0.757 | 170 | 240 | 0.857 | 0.905 | 1.41 |
| 0.3 | 5 | 0.85 | 0.786 | 120 | 190 | 0.838 | 0.890 | 1.58 |
| 0.3 | 5 | 0.90 | 0.812 | 90 | 160 | 0.775 | 0.875 | 1.78 |
| 0.3 | 10 | 0.70 | 0.683 | 820 | 990 | 0.892 | 0.905 | 1.21 |
| 0.3 | 10 | 0.75 | 0.720 | 510 | 670 | 0.882 | 0.906 | 1.31 |
| 0.3 | 10 | 0.80 | 0.752 | 340 | 500 | 0.858 | 0.899 | 1.47 |

| $\phi$ | $p$ | $C$ | $C_{\text{adj}}$ | $n_{\text{orig}}$ | $n_{\text{adj}}$ | $Med(s_{n_{\text{orig}}})$ | $Med(s_{n_{\text{adj}}})$ | $\frac{n_{\text{adj}}}{n_{\text{orig}}}$ |
|---|---|---|---|---|---|---|---|---|
| 0.3 | 10 | 0.85 | 0.780 | 240 | 400 | 0.827 | 0.891 | 1.67 |
| 0.3 | 10 | 0.90 | 0.805 | 170 | 330 | 0.773 | 0.879 | 1.94 |
| 0.3 | 15 | 0.70 | 0.683 | 1,240 | 1,480 | 0.884 | 0.901 | 1.19 |
| 0.3 | 15 | 0.75 | 0.719 | 770 | 1,020 | 0.873 | 0.903 | 1.32 |
| 0.3 | 15 | 0.80 | 0.751 | 520 | 760 | 0.853 | 0.898 | 1.46 |
| 0.3 | 15 | 0.85 | 0.779 | 360 | 610 | 0.818 | 0.892 | 1.69 |
| 0.3 | 15 | 0.90 | 0.803 | 260 | 510 | 0.767 | 0.876 | 1.96 |
| 0.3 | 20 | 0.70 | 0.683 | 1,650 | 1,990 | 0.885 | 0.903 | 1.21 |
| 0.3 | 20 | 0.75 | 0.719 | 1,030 | 1,360 | 0.874 | 0.900 | 1.32 |
| 0.3 | 20 | 0.80 | 0.750 | 690 | 1,030 | 0.849 | 0.896 | 1.49 |
| 0.3 | 20 | 0.85 | 0.778 | 480 | 820 | 0.819 | 0.891 | 1.71 |
| 0.3 | 20 | 0.90 | 0.801 | 340 | 680 | 0.763 | 0.875 | 2.00 |
| 0.5 | 5 | 0.70 | 0.683 | 350 | 420 | 0.900 | 0.910 | 1.20 |
| 0.5 | 5 | 0.75 | 0.720 | 220 | 280 | 0.888 | 0.911 | 1.27 |
| 0.5 | 5 | 0.80 | 0.752 | 150 | 210 | 0.855 | 0.897 | 1.40 |
| 0.5 | 5 | 0.85 | 0.781 | 100 | 170 | 0.822 | 0.898 | 1.70 |
| 0.5 | 5 | 0.90 | 0.806 | 70 | 140 | 0.758 | 0.873 | 2.00 |
| 0.5 | 10 | 0.70 | 0.682 | 700 | 850 | 0.883 | 0.905 | 1.21 |
| 0.5 | 10 | 0.75 | 0.717 | 440 | 590 | 0.868 | 0.908 | 1.34 |
| 0.5 | 10 | 0.80 | 0.748 | 290 | 440 | 0.855 | 0.897 | 1.52 |
| 0.5 | 10 | 0.85 | 0.775 | 210 | 360 | 0.815 | 0.892 | 1.71 |
| 0.5 | 10 | 0.90 | 0.798 | 150 | 300 | 0.756 | 0.878 | 2.00 |
| 0.5 | 15 | 0.70 | 0.681 | 1,040 | 1,270 | 0.883 | 0.904 | 1.22 |
| 0.5 | 15 | 0.75 | 0.716 | 650 | 880 | 0.873 | 0.897 | 1.35 |
| 0.5 | 15 | 0.80 | 0.747 | 440 | 670 | 0.851 | 0.898 | 1.52 |
| 0.5 | 15 | 0.85 | 0.773 | 310 | 540 | 0.818 | 0.889 | 1.74 |
| 0.5 | 15 | 0.90 | 0.795 | 220 | 450 | 0.757 | 0.877 | 2.05 |
| 0.5 | 20 | 0.70 | 0.681 | 1,390 | 1,710 | 0.879 | 0.904 | 1.23 |
| 0.5 | 20 | 0.75 | 0.715 | 870 | 1,200 | 0.870 | 0.904 | 1.38 |
| 0.5 | 20 | 0.80 | 0.746 | 580 | 900 | 0.849 | 0.898 | 1.55 |
| 0.5 | 20 | 0.85 | 0.772 | 410 | 730 | 0.817 | 0.896 | 1.78 |
| 0.5 | 20 | 0.90 | 0.794 | 290 | 610 | 0.757 | 0.881 | 2.10 |

*Figure S3. Sample size required to ensure* $PrAP(\hat{s}_n) = 0.8$ *using either i) the simulation-based approach (blue) or ii) the analytical approach (purple) for varying values of the C-statistic. B. Probability of acceptable calibration for the sample sizes based on i) and ii). For all combinations of the C-statistic (range 0.65-0.9), outcome prevalence (0.1, 0.3, 0.5) and number of predictors (5, 10, 20) the sample size was calculated assuming MLE. Adjusted C-values were used for step a) and step c) of the algorithm for the analytical calculation for* $C \geq 0.75$.

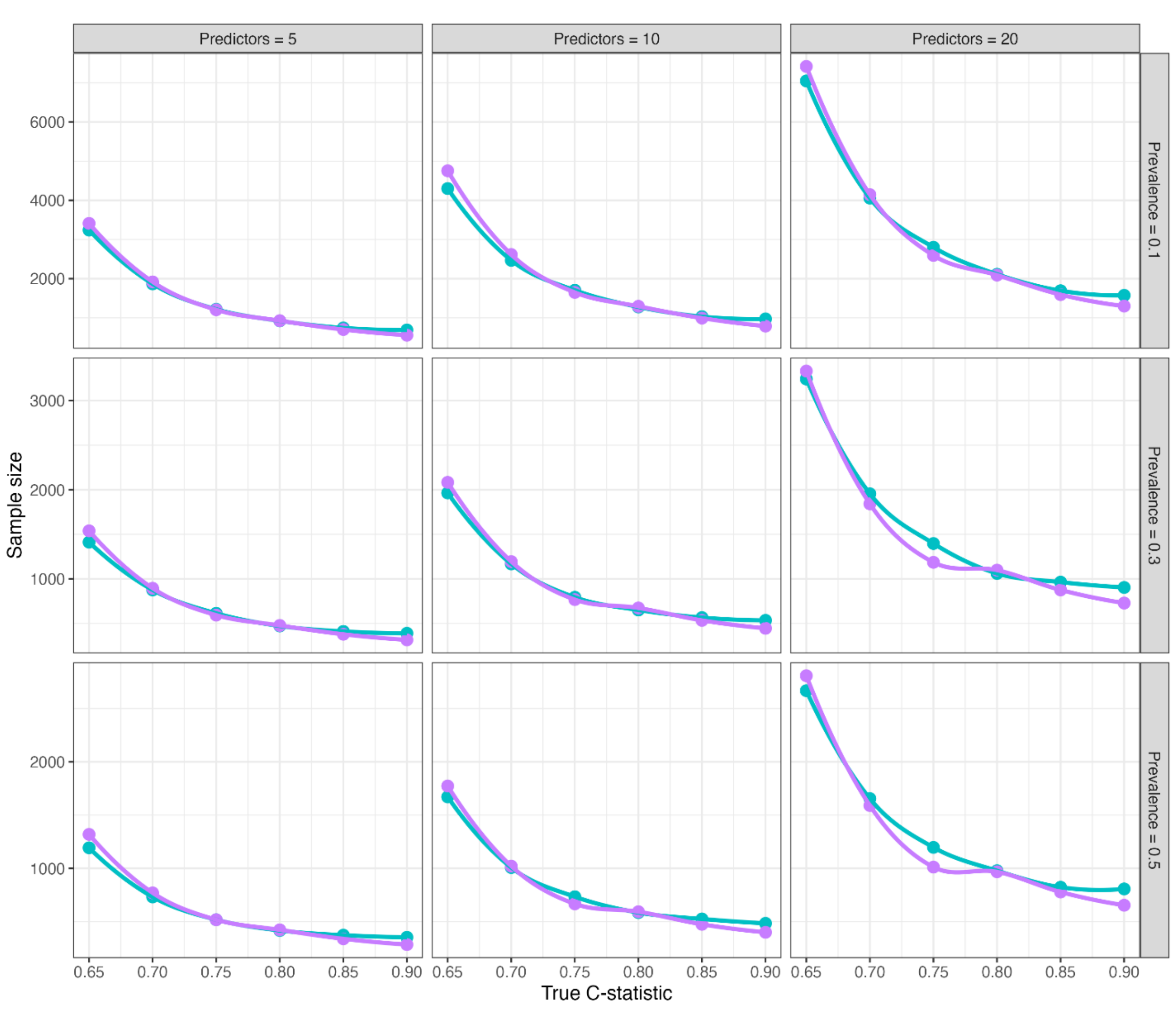

*Table S2.* $\sqrt{Var(\hat{s}_n)}$ *and* $\sqrt{Var_{app}(\hat{s}_n)}$ *and when sample size varies (500-4000) and the C-statistic and prevalence are fixed to 0.7 and 0.1, respectively.*

| n | $\mathbb{E}(\hat{s}_n)$ | $\mathrm{PrAP}(\hat{s}_n)$ | $\sqrt{\mathrm{Var}(\hat{s}_n)}$ | $\sqrt{\mathrm{Var}_{\mathrm{app}}(\hat{s}_n)}$ |
|---|---|---|---|---|
| **4,000** | **0.953** | **0.93** | **0.07070** | **0.0698** |
| **3,000** | **0.939** | **0.85** | **0.08018** | **0.0816** |
| **2,000** | **0.911** | **0.72** | **0.09504** | **0.0943** |
| 1,500 | 0.886 | 0.59 | 0.10683 | 0.1044 |
| 1,000 | 0.830 | 0.38 | 0.12175 | 0.1176 |
| 800 | 0.794 | 0.28 | 0.12996 | 0.1251 |
| 700 | 0.770 | 0.23 | 0.13466 | 0.1272 |
| 500 | 0.700 | 0.12 | 0.14458 | 0.1333 |

*Table S3. The standard deviation of the calibration slope (via simulation versus analytical) for a combination of c-statistic, prevalence, and number of predictor values. For all combinations of the C-statistic (range 0.65-0.9), outcome prevalence (0.1, 0.3, 0.5) and number of predictors (5, 10, 20) the sample size was calculated assuming MLE. The sample size was chosen by simulation to ensure that* $PrAP(\hat{s}_n) = 0.8$. *The adjusted C value was used in the approximate variance formula for* $C \geq 0.75$.

| $c$ | $p$ | ϕ | Sample size | $\mathbb{E}_{\text{sim}}(\hat{s}_n)$ | $\text{SD}_{\text{sim}}(\hat{s}_n)$ | $\text{SD}_{\text{app}}(\hat{s}_n)$ | $\frac{\text{SD}_{\text{sim}}(\hat{s}_n)}{\text{SD}_{\text{app}}(\hat{s}_n)}$ |
|---|---|---|---|---|---|---|---|
| 0.65 | 5 | 0.1 | 3,240 | 0.951 | 0.1090 | 0.1049 | 1.039 |
| 0.65 | 5 | 0.3 | 1,397 | 0.952 | 0.1117 | 0.1082 | 1.033 |
| 0.65 | 5 | 0.5 | 1,157 | 0.949 | 0.1080 | 0.1097 | 0.984 |
| 0.65 | 10 | 0.1 | 4,184 | 0.921 | 0.0875 | 0.0894 | 0.979 |
| 0.65 | 10 | 0.3 | 1,933 | 0.921 | 0.0881 | 0.0890 | 0.990 |
| 0.65 | 10 | 0.5 | 1,615 | 0.918 | 0.0883 | 0.0899 | 0.983 |
| 0.65 | 20 | 0.1 | 7,245 | 0.903 | 0.0646 | 0.0666 | 0.970 |
| 0.65 | 20 | 0.3 | 3,148 | 0.903 | 0.0678 | 0.0684 | 0.992 |
| 0.65 | 20 | 0.5 | 2,784 | 0.902 | 0.0651 | 0.0672 | 0.968 |
| 0.70 | 5 | 0.1 | 1,779 | 0.953 | 0.1104 | 0.1065 | 1.037 |
| 0.70 | 5 | 0.3 | 877 | 0.946 | 0.1074 | 0.1043 | 1.030 |
| 0.70 | 5 | 0.5 | 718 | 0.951 | 0.1079 | 0.1081 | 0.998 |
| 0.70 | 10 | 0.1 | 2,608 | 0.924 | 0.0856 | 0.0853 | 1.004 |
| 0.70 | 10 | 0.3 | 1,147 | 0.919 | 0.0891 | 0.0886 | 1.006 |
| 0.70 | 10 | 0.5 | 1,005 | 0.924 | 0.0874 | 0.0888 | 0.985 |
| 0.70 | 20 | 0.1 | 3,991 | 0.899 | 0.0671 | 0.0671 | 1.001 |
| 0.70 | 20 | 0.3 | 2,002 | 0.904 | 0.0658 | 0.0660 | 0.997 |
| 0.70 | 20 | 0.5 | 1,646 | 0.900 | 0.0654 | 0.0676 | 0.968 |
| 0.75 | 5 | 0.1 | 1,174 | 0.949 | 0.1088 | 0.1063 | 1.023 |
| 0.75 | 5 | 0.3 | 593 | 0.947 | 0.1079 | 0.1053 | 1.025 |
| 0.75 | 5 | 0.5 | 539 | 0.952 | 0.1079 | 0.1040 | 1.038 |
| 0.75 | 10 | 0.1 | 1,614 | 0.921 | 0.0891 | 0.0880 | 1.013 |
| 0.75 | 10 | 0.3 | 824 | 0.923 | 0.0874 | 0.0870 | 1.004 |
| 0.75 | 10 | 0.5 | 742 | 0.923 | 0.0864 | 0.0859 | 1.006 |
| 0.75 | 20 | 0.1 | 2,835 | 0.906 | 0.0654 | 0.0653 | 1.001 |
| 0.75 | 20 | 0.3 | 1,402 | 0.902 | 0.0650 | 0.0652 | 0.997 |
| 0.75 | 20 | 0.5 | 1,190 | 0.905 | 0.0659 | 0.0665 | 0.991 |
| 0.80 | 5 | 0.1 | 886 | 0.955 | 0.1088 | 0.1037 | 1.049 |
| 0.80 | 5 | 0.3 | 484 | 0.950 | 0.1043 | 0.1015 | 1.027 |

| $c$ | $p$ | ϕ | Sample size | $\mathbb{E}_{\text{sim}}(\hat{s}_n)$ | $\text{SD}_{\text{sim}}(\hat{s}_n)$ | $\text{SD}_{\text{app}}(\hat{s}_n)$ | $\frac{\text{SD}_{\text{sim}}(\hat{s}_n)}{\text{SD}_{\text{app}}(\hat{s}_n)}$ |
|---|---|---|---|---|---|---|---|
| 0.80 | 5 | 0.5 | 413 | 0.946 | 0.1050 | 0.1029 | 1.020 |
| 0.80 | 10 | 0.1 | 1,261 | 0.920 | 0.0863 | 0.0838 | 1.030 |
| 0.80 | 10 | 0.3 | 651 | 0.923 | 0.0889 | 0.0850 | 1.045 |
| 0.80 | 10 | 0.5 | 582 | 0.920 | 0.0879 | 0.0843 | 1.043 |
| 0.80 | 20 | 0.1 | 2,087 | 0.903 | 0.0649 | 0.0639 | 1.016 |
| 0.80 | 20 | 0.3 | 1,080 | 0.901 | 0.0656 | 0.0644 | 1.018 |
| 0.80 | 20 | 0.5 | 953 | 0.904 | 0.0658 | 0.0647 | 1.017 |
| 0.85 | 5 | 0.1 | 735 | 0.948 | 0.1101 | 0.0985 | 1.117 |
| 0.85 | 5 | 0.3 | 410 | 0.946 | 0.1037 | 0.0982 | 1.056 |
| 0.85 | 5 | 0.5 | 359 | 0.948 | 0.1043 | 0.0997 | 1.047 |
| 0.85 | 10 | 0.1 | 1,048 | 0.924 | 0.0870 | 0.0804 | 1.082 |
| 0.85 | 10 | 0.3 | 568 | 0.924 | 0.0879 | 0.0814 | 1.079 |
| 0.85 | 10 | 0.5 | 505 | 0.926 | 0.0881 | 0.0820 | 1.074 |
| 0.85 | 20 | 0.1 | 1,719 | 0.902 | 0.0666 | 0.0613 | 1.086 |
| 0.85 | 20 | 0.3 | 920 | 0.904 | 0.0675 | 0.0626 | 1.078 |
| 0.85 | 20 | 0.5 | 828 | 0.903 | 0.0656 | 0.0625 | 1.050 |
| 0.90 | 5 | 0.1 | 685 | 0.948 | 0.1057 | 0.0910 | 1.161 |
| 0.90 | 5 | 0.3 | 393 | 0.949 | 0.1036 | 0.0923 | 1.123 |
| 0.90 | 5 | 0.5 | 348 | 0.949 | 0.1057 | 0.0934 | 1.131 |
| 0.90 | 10 | 0.1 | 975 | 0.923 | 0.0884 | 0.0743 | 1.190 |
| 0.90 | 10 | 0.3 | 537 | 0.920 | 0.0869 | 0.0765 | 1.136 |
| 0.90 | 10 | 0.5 | 472 | 0.921 | 0.0885 | 0.0778 | 1.137 |
| 0.90 | 20 | 0.1 | 1,561 | 0.907 | 0.0662 | 0.0577 | 1.148 |
| 0.90 | 20 | 0.3 | 893 | 0.904 | 0.0678 | 0.0583 | 1.164 |
| 0.90 | 20 | 0.5 | 802 | 0.906 | 0.0671 | 0.0587 | 1.143 |

## 2 Simulation Design

### Aims

To obtain the sampling distribution of measures of predictive performance when adhering to the sample size requirements for model development to limit model overfitting, to achieve either a) target expected performance or b) a target probability of acceptable performance.

### Data generating mechanism

We aim to generate $n_{sim}$ development and validation datasets with binary outcomes from the following logistic regression model

$$\text{logit}\big(P(Y = 1|\boldsymbol{X})\big) = \eta = \beta_0 + \boldsymbol{X}^\top \boldsymbol{\beta_1}$$

where $Y$ is the binary outcome, $\boldsymbol{\beta_1} = \big(\beta_1, \dots \beta_p\big)^\top$ is a $p$-dimensional vector of regression coefficients and $\boldsymbol{X} = \big(X_1, \dots X_p\big)^\top$ is a vector of covariate values. The values of regression coefficients are chosen to correspond to an assumed (input) prevalence, $\phi$ and a C-statistic, $C$.

In the most general situation , $\boldsymbol{X}$ will have an arbitrary distribution which is user defined/generated. For example, $\boldsymbol{X}$ can be generated to mimic the empirical distribution of $\boldsymbol{X}$ from an existing dataset from where we also have assumed parameter values $\beta_0^*$ and $\boldsymbol{\beta}_1^*$. On this occasion, we can use simulation and optimisation to obtain suitable values $a_0$ and $f$ such $\beta_0 = \beta_0^* + a_0$ and $\boldsymbol{\beta}_1 = f\boldsymbol{\beta}_1^*$ to ensure that the overall prevalence is $\phi$ and C-statistic, $C$ meet some target values for the assumed true model.

Without loss of generality, and unless otherwise stated, we assume that $\boldsymbol{X} \sim MVN\big(\boldsymbol{0}, \boldsymbol{I}_p\big)$ and $\beta_{1j} = \beta \;\; \forall\, j = 1, \dots p$. This suggests that $\eta \sim N(\mu, \sigma^2)$ where $\sigma^2 = \sum_j \beta_{1j}^2 = p\beta^2$. This enables us to calculate $\mu$ and $\beta$ using analytical formulae or numerical integration/simulation and optimisation. For the more general case where $\boldsymbol{X}$ has a generic distribution (e.g. based on existing data) $\beta_0$ and $\beta_{1j}$ are set such that the true model corresponds to the target $C$ and prevalence.

The simulation process for given values of $p$, $\phi$ and $C$ and sample size, $n$, is as follows:

1. Generate a training dataset of size $n$ with the following steps
   a. Generate $\boldsymbol{X_i}$
   b. Calculate $\eta_i = \beta_0 + \boldsymbol{X}_i^\top \boldsymbol{\beta_1}$
   c. Calculate $\pi_i = logit^{-1}(\eta_i)$
   d. Generate $Y_i \sim Bernoulli(\pi_i)$
2. Generate a validation dataset with size $n_{val}$ using steps 1a-1d. The validation dataset should be large enough such that performance measures can be estimated with small variability.
3. Fit the model on the training dataset using a statistical method to estimate $\widehat{\boldsymbol{\beta}}_{\boldsymbol{n}}$. The default fitting method is MLE, while other methods (e.g. penalised regression) are also possible.

4. Validate the model on the validation dataset and calculate a predictive performance measure, $\theta$.
5. Repeat steps 1-6 $n_{sim}$ times such that the Monte Carlo simulation error is sufficiently small; $n_{sim} = 1000 - 2000$ will be an appropriate number in most scenarios.

**Performance measures and targets**

We let $\hat{\theta}_{n,j},\ j = 1, \dots, n_{\text{sim}}$ denote the calculated predictive performance measure for the j-*th* development dataset of size $n$. For example, $\theta$ can be the calibration slope, C-statistic, Brier Score, MAPE, Net Benefit etc.

Possible targets of interest are $\mathbb{E}(\hat{\theta}_n)$ or $\text{Med}(\hat{\theta}_n)$, $\text{Var}(\hat{\theta}_n)$ and $\text{PrAP}(\hat{\theta}_n) = P(l_\theta \leq \hat{\theta}_n \leq u_\theta)$ where

1) $\mathbb{E}(\hat{\theta}_n) = \bar{\theta}_n = \frac{\sum_{j=1}^{n_{\text{sim}}} \hat{\theta}_{n,j}}{n_{\text{sim}}}$ and $\text{Med}(\hat{\theta}_n) = \tilde{\theta} = \text{Median}\{\hat{\theta}_{n,1}, \hat{\theta}_{n,2}, \dots, \hat{\theta}_{n,n_{\text{sim}}}\}$
2) $\text{Var}(\hat{\theta}_n) = \frac{\sum_{j=1}^{n_{\text{sim}}} (\hat{\theta}_{n,j} - \bar{\theta}_n)^2}{n_{\text{sim}} - 1}$
3) $\text{PrAP}(\hat{\theta}_n) = \frac{\sum_{j=1}^{n_{\text{sim}}} I(\hat{\theta}_{n,j} \in [l_\theta, u_\theta])}{n_{\text{sim}}}$

For the simulation studies presented in this paper, at least two 'baseline' sample sizes are of potential interest (assuming that MLE is used to fit the model).

a) The sample size that corresponds $\text{Med}(\hat{s}_n) = 0.9$

b) The sample size that corresponds to $\text{PrAP}(\hat{s}_n) = \Pr(0.85 \leq \hat{s}_n \leq 1.15) = 0.8$

By definition, $\text{Med}(\hat{s}_n) = 0.9$ for the first, and $\text{PrAP}(\hat{s}_n) = 0.8$ for the second. Approximate sample size estimators for a) and b) were discussed in Sections 3 and 4 of the main paper.

## 3 Obtaining the adjusted C for bias reduction in the calibration sample size formula

Let $Y$ and $\boldsymbol{X}$ denote a binary outcome and a p-dimensional vector of continuous covariates, respectively. Let us consider the following logistic regression model for $Y|\boldsymbol{X}$:

$$\text{logit}(P(Y|\boldsymbol{X})) = \beta_0 + \boldsymbol{X}^\top \boldsymbol{\beta}_1 \quad (*).$$

Suppose that the true data generating mechanism (DGM) follows from a Linear Discriminant Analysis (LDA) representation for $\boldsymbol{X}|Y$:

$$\boldsymbol{X}|Y = k \sim MVN(\boldsymbol{\mu}_k, \boldsymbol{\Sigma}), k = 0,1$$

where $\boldsymbol{\mu}_k \in \mathbb{R}^p$ are the conditional means for $\boldsymbol{X}|Y = k$, and $\boldsymbol{\Sigma}$ is the common variance-covariance matrix. This data-generating mechanism corresponds to a 'retrospective sampling scheme'. For this model, the log-odds ratios are $\boldsymbol{\delta_1} = \boldsymbol{\Sigma}^{-1}(\boldsymbol{\mu}_1 - \boldsymbol{\mu}_0)$. When the assumptions of the LDA model (conditional normality and homogeneity of variance) hold, as occurs when the DGM follows the retrospective design, then the log-odds ratios are asymptotically equal under the logistic and LDA representations with $\widehat{\boldsymbol{\beta}}_1^{LR} = \widehat{\boldsymbol{\delta}}_1$.

A more commonly assumed DGM is based on a logistic regression model for $P(Y|\boldsymbol{X})$. This corresponds to a 'prospective sampling scheme'. Let us assume that the true DGM is

$$\operatorname{logit}\big(P(Y|\boldsymbol{X})\big) = \beta_0 + \boldsymbol{X}^\top \boldsymbol{\beta}_1 \text{ with } \boldsymbol{X} \sim MVN(\boldsymbol{\mu}, \boldsymbol{\Sigma}) \quad (**).$$

and let $C$ denote the C-statistic that corresponds to model $(**)$. Under the prospective DGM, the equivalence between the estimated log-odds ratios based on the LDA model and $\boldsymbol{\beta}_1$ no longer holds, in general. This is because marginal normality of $\boldsymbol{X}$ does not guarantee conditional normality of $\boldsymbol{X}|Y$ and, hence, the assumptions of the LDA model may be violated. In particular, when $C$ is relatively low (e.g. $C \leq 0.75$) marginal normality of $\boldsymbol{X}$ also corresponds approximately to conditional normality for $\boldsymbol{X}|Y$. Then, $\widehat{\boldsymbol{\delta}} \approx \widehat{\boldsymbol{\beta}}_1$. When discrimination is high, marginal normality for $\boldsymbol{X}$ does not translate into conditional normality for $\boldsymbol{X}|Y$. The LDA assumptions are violated, and consequently $\widehat{\boldsymbol{\delta}} \neq \widehat{\boldsymbol{\beta}}_{\boldsymbol{1}}$. This corresponds to the range of scenarios where van Houwelingen's shrinkage (calibration slope) approximation in the main manuscript does not work.

The full algorithm to compute $C_{\text{adj}}$ as seen in the main paper is as follows:

a) Consider a normal linear predictor $\eta \sim N(\mu, \sigma^2)$ that corresponds to the required $C$ and prevalence e.g. using the approach of Pavlou (2024)[1].
b) Express the linear predictor corresponding to model $(*)$ as a linear combination of independent standard normal predictor variables $X_j \sim N(0,1)$ and regression coefficients, $\beta_j$. Without loss of generality, assume $\beta_j = \beta_k = \beta, j = 1, \ldots p.$. Then, $\sigma^2 = \operatorname{var}(\beta_0 + \boldsymbol{X}^\top \boldsymbol{\beta}_1) = p\beta^2$ which gives $\beta = \sqrt{\sigma^2/p}$ and $\beta_0 = \mu$.
c) Simulate a large dataset (e.g. size 1,000,000) of $\boldsymbol{X} \sim MVN(\boldsymbol{0}, I_p)$ and outcomes from the (prospective DGM) model
$$\operatorname{logit}\big(P(Y|\boldsymbol{X})\big) = \beta_0 + \boldsymbol{X}^\top \boldsymbol{\beta}_1.$$
d) Calculate the corresponding log-odds ratio, $\hat{\delta}_j$ of each covariate assuming an LDA model (as the sample size is very large, we subsequently use $\delta_j$ instead of $\hat{\delta}_j$ ):
$$\delta_j = (\bar{X}_{j1} - \bar{X}_{j0})/\sigma_j^2 \; j = 1, \ldots p$$
where $\bar{X}_{j1} = \sum X_{ij1}/n_1$ and $\bar{X}_{j0} = \sum X_{ij0}/n_0$ are the conditional means for $Y = 0$ and $Y = 1$, respectively, and $\sigma_j^2 = \frac{(n_0-1)\,\sigma_{j0}^2 + (n_1-1)\,\sigma_{j1}^2}{n_0+n_1-2}$ is the pooled variance.
e) Compute the adjusted linear predictor based on the LDA log-odds ratios: $\eta_{\text{adj}} = \delta_0 + \sum \delta_j X_j$.
f) Compute the adjusted C-statistic, $C_{\text{adj}}$.

When $C$ is relatively small, e.g. $C \leq 0.7$, $\beta_j \approx \delta_j$ and hence $C_{\text{adj}} \approx C$. However, for larger values of $C$, $\delta_j < \beta_j$, which leads to $C_{\text{adj}} < C$. We assessed the effect of this correction for a range of values for $C$ (0.65-0.9), $p$ (5, 10, 20) and $\phi$ $(0.1, 0.3, 0.5)$. The results are in Table S1. The degree of deviation between $C_{\text{adj}}$ and $C$ is consistent with the degree of bias seen previously in the calibration formula for $\mathbb{E}(\hat{s}_n)$. Use of $C_{\text{adj}}$ leads to substantial bias reduction in the required sample size, with the expected calibration slope being very close to the nominal value of 0.9 for all scenarios.

## 4 Empirical approximation for the variance of the calibration slope

As before, we let $Y$ be a binary outcome and $\boldsymbol{X}$ a $p$-dimensional vector of covariate values. We assume that $Y$ is modelled in terms of $\boldsymbol{X}$ through the following logistic regression model

$$\text{logit}\big(P(Y|\boldsymbol{X})\big) = \eta$$

where $\eta = \beta_0 + \boldsymbol{X}^\top \boldsymbol{\beta}_1$ and $\boldsymbol{\beta} = (\beta_0, \boldsymbol{\beta}_1^\top)^\top$ is a vector of regression coefficients. For a specified distribution of $\boldsymbol{X}$ we assume that the value of $\boldsymbol{\beta}$ corresponds to prevalence $\phi$ and C-statistic $C$, with $\mathbb{E}(\eta) = \beta_0$ and $\text{Var}(\eta) = \sigma_\eta^2$.

We are interested in the variability of the calibration slope in two scenarios.

A. *Finite development data* of size $n$. We assume that for a given (finite) development sample size, $n$, for given $p$, $\phi$ and $C$, the model above will be overfitted, with expected calibration slope (in validation data from the same population), $\mathbb{E}(\hat{s}_n) = S_n$. The sampling distribution of the calibration slope is obtained by repeatedly fitting the model above on training datasets of size $n$ and obtaining the calibration slope $\hat{s}_n$ on a *very large validation dataset* via the model

$$\text{logit}\big(P(Y_{\text{val}} = 1|\boldsymbol{X}_{\textbf{val}})\big) = a_0 + s_n \times \hat{\eta}_n \,, \qquad (1)$$

where $\hat{\eta}_n = \hat{\beta}_{0n} + \boldsymbol{X}_{\textbf{val}}^\top \widehat{\boldsymbol{\beta}}_{1n}$ and $\widehat{\boldsymbol{\beta}}_n$ denotes an estimate of $\boldsymbol{\beta}$ in a development sample of size $n$. Over repeated sampling of development samples of size $n$, the expected calibration slope is $S_n = \mathbb{E}(\hat{s}_n)$. The variability in $\hat{s}_n$, $\text{Var}(\hat{s}_n)$, is due to the limited size of the development dataset(s). As $n \to \infty$, $\mathbb{E}(\hat{s}_n) \to 1$ and $\text{Var}(\hat{s}_n) \to 0$.

B. *Finite Validation data*. We assume that for given $p$, $\phi$ and $C$, an overfitted model *with known expected calibration slope* $S_n$ is validated on validation datasets of size $m$. This is the setting considered in Pavlou et al. (2021)[2]. The sampling distribution of the estimated calibration slope is obtained by repeatedly estimating the calibration slope on finite validation datasets via the model:

$$\text{logit}\Big(P\big(Y_{\text{val}} = 1\big|\boldsymbol{X}_{\text{val,m}}\big)\Big) = a + s_{n,m} \times \eta_n \,, \qquad (2)$$

where $\eta_n = \frac{1}{S_n}\big(\beta_0 + \boldsymbol{X}_{\text{val},m}^\top\, \boldsymbol{\beta}\big)$ and $\boldsymbol{\beta}$ are fixed at their true values.

In scenario A, $\hat{s}_n$ is calculated on a very large validation dataset and hence the estimation variability is negligible. The variability in $\hat{s}_n$ *across* development samples is solely due to variability in the estimation of $\boldsymbol{\beta}$. Asymptotically, $\widehat{\boldsymbol{\beta}}_n \sim \text{N}\Big(\boldsymbol{\beta}, \frac{1}{n} I^{-1}(\boldsymbol{\beta})\Big)$, where $I(\boldsymbol{\beta})$ denotes Fisher's information for $\boldsymbol{\beta}$. The uncertainty stemming from the estimation of $\hat{\beta}_n$ propagates into the uncertainty around $\hat{s}_n$. In scenario B, the validation datasets (and $\boldsymbol{X}_{\text{val,m}}$) are finite with size $m$. With the other components of the fitted linear predictor ($\boldsymbol{\beta}$ and $S_n$) fixed, the variability in $\hat{s}_{n,m}$ across validation datasets is solely due to variability in the estimation of $s_{n,m}$: $\hat{s}_{n,m} \sim \mathrm{N}\left(S_n, \frac{1}{m} I^{-1}(S_n)\right)$.

When (i) the development datasets considered in model (1) are of the same size $n$ as the validation datasets considered in model (2), and (ii) $n$ is sufficiently large, the uncertainty in $\hat{s}_n$ attributable to estimating $\boldsymbol{\beta}$ from $n$ observations will be very similar to the expected variability in $\hat{s}_{n,n}$, $\mathbb{E}\left[\mathrm{Var}\left(\hat{s}_{n,n} | S_n\right)\right]$, attributable to estimating $s$ from $n$ observations when the linear predictor is known, i.e. $\mathrm{Var}(\hat{s}_n) \approx \mathbb{E}\left(\mathrm{Var}\left(\hat{s}_{n,n} | S_n\right)\right)$. This similarity holds because both sources of error arise from parameter estimation with $n$ observations in a logistic regression setting.

Assuming a linear discriminant analysis model for (2) and assuming conditional normality with equal variances per class, the following formula for the expected variability for $\mathbb{E}\left[\mathrm{Var}\left(\hat{s}_{n,n} | S_n\right)\right]$ was proposed by Pavlou et al. (2021)[2]:

$$\mathbb{E}\left[\mathrm{Var}\left(\hat{s}_{n,n} | S_n\right)\right] \approx \frac{S_n^2}{2\,\phi\,(1-\phi)\,n\,\Phi^{-1}(C)^2} + \frac{2\,S_n^2}{n-2} \qquad (3).$$

We note that approximation (3) was seen to be slightly biased for $C \geq 0.75$ [2], for the same reasons Riley's[3] calibration equation for $\mathbb{E}(\hat{s}_n)$ is biased for high $C$. Hence, the adjustment to the input value of $C$ discussed in Section 4 of the main paper is applied for the variance formula (with $p = 1$ predictor, corresponding to the linear predictor in model (2) above).

Under this approximation, $\mathrm{Var}(\hat{s}_n)$ can be approximated by $\mathrm{Var}_{\mathrm{app}}(\hat{s}_n) \approx \frac{S_n^2}{2\,\phi\,(1-\phi)\,n\,\Phi^{-1}(C)^2} + \frac{2\,S_n^2}{n-2}$.

This empirical approximation is supported by simulation results.

To quantify the effect of sample size on the similarity between A and B we considered the simulation scenario of Section 5.2 with $p = 10$, $C = 0.7$ and $\phi = 0.1$. We then used simulation to calculate $\mathrm{Var}(\hat{s}_n)$ and $\mathrm{Var}_{\mathrm{app}}(\hat{s}_n)$ for $n = 500$ to $4000$. As shown in Table S2, the contributions from the two sources of variability are very similar for sample sizes for which we intend to use $\mathrm{Var}_{\mathrm{app}}(\hat{s}_n)$ for sample size calculations, i.e. around $\mathrm{PrAP}(\hat{s}_n) = 0.8$. Specifically, $\sqrt{\mathrm{Var}(\hat{s}_n)} / \sqrt{\mathrm{Var}_{\mathrm{app}}(\hat{s}_n)}$ is between 0.99 and 1.02 for sample sizes between 1500 (for which $\mathrm{PrAP}(\hat{s}_n) = 0.6$) to 4000 (for which $\mathrm{PrAP}(\hat{s}_n) = 0.95$). Table S3 further demonstrates the agreement between $\mathrm{Var}(\hat{s}_n)$ and $\mathrm{Var}_{\mathrm{app}}(\hat{s}_n)$ for a wide range of outcome prevalence, number of predictors and C-statistic values at sample sizes, $n$, for which $\mathrm{PrAP}(\hat{s}_n) = 0.8$; these are sample sizes around which we intend to use $\mathrm{Var}_{\mathrm{app}}(\hat{s}_n)$ for sample size calculations.

**References for the Supplement**